\documentclass[conference]{IEEEtran}
\IEEEoverridecommandlockouts

\usepackage{verbatim}
\usepackage[
backend=biber,
style=numeric,
sorting=ynt,
]{biblatex}
\nocite{*}
\usepackage{glossaries}
\usepackage{amsmath,amssymb,amsfonts}
\usepackage[english]{babel}
\usepackage[utf8]{inputenc}
\usepackage{algorithm}
\usepackage[noend]{algpseudocode}
\usepackage{graphicx}
\usepackage{subfigure}
\usepackage{textcomp}
\usepackage{csquotes}
\usepackage{xcolor}
\def\BibTeX{{\rm B\kern-.05em{\sc i\kern-.025em b}\kern-.08em
    T\kern-.1667em\lower.7ex\hbox{E}\kern-.125emX}}
\begin{document}

\title{
Bitcoin P2P Network Measurements: A testbed study of the effect of peer selection on transaction propagation and confirmation times

}

\author{\IEEEauthorblockN{Befekadu G. Gebraselase, Bjarne E. Helvik, Yuming Jiang}
\IEEEauthorblockA{\textit{Department of Information Security and Communication Technology} \\
\textit{NTNU, Norwegian University of Science and Technology, Trondheim, Norway}\\
\{befekadu.gebraselase, bjarne, yuming.jiang\}@ntnu.no}

}

\maketitle

\begin{abstract}
 Bitcoin is the first and the most extensive decentralized electronic cryptocurrency system that uses blockchain technology. It uses a peer-to-peer (P2P) network to operate without a central authority and propagate system information such as transactions or blockchain updates. The communication between participating nodes is highly relying on the underlying network infrastructure to facilitate a platform. Understanding the impact of peer formation strategies, peer list, and delay are vital on understanding node to node communication. To this aim, we performed an extensive study on the transaction characteristic of Bitcoin through a Testbed. The analysis shows that peer selection strategies affect the transactions propagation and confirmation time. Moreover, the default distance-based peer selection strategy in Bitcoin performs less when there is high arrival intensity and creates high number forks. 
\end{abstract}

\section{Introduction}
Bitcoin is the first well-known decentralized electronic P2P system that uses blockchain technology.  It adapts a cryptographic proof of work (PoW) mechanism that allows anonymous peers to create and validate transactions through the underlying peer-to-peer (P2P) network~\cite{nakamoto2008bitcoin}. The P2P network is vital to the communications of the blockchain system~\cite{LocalApproch}\cite{p2pdiscovery}.  The nodes send and receive messages via the underlying network infrastructure while the P2P topology is formed at the application layer~\cite{discovery}.  The way nodes form an overlay topology affects the overall performance, such as transactions confirmation time~\cite{trasactionConfirmation}, block and transaction propagation delay~\cite{prop}\cite{transHandling}, fork rate~\cite{prop}, and stability of the ledger. To this aim, we prepared a testbed to analyze the impact of peer-to-peer topology formation, end-to-end delay, and bandwidth limitation on the performance of Bitcoin. 

Bitcoin operates to distribute the ledger among all the participants in a flooding P2P network~\cite{p2pnetwork}. When a node tries to join the Bitcoin network, it uses a hardcoded seed to reach out to the nodes nearby. Through getaddress and node discovery, each node updates/creates eight peers by default (outgoing connection), but it can have 125 inbound connections. The logical connection between participating nodes creates a dense P2P overlay topology, a mesh network~\cite{p2ptoplogy}\cite{p2pdiscovery}. This P2P topology is responsible for broadcasting new updates to peers by which they learn and inform each other about transactions and blocks~\cite{p2pdiscovery}. The reachability of these messages affects the ability of the system to process more transactions and secure the interactions ~\cite{LocalApproch}\cite{p2pdiscovery}. 


In Bitcoin, the average inter-block generation time is 10 minutes. This enables all the newly generated blocks to reach the maximum number of nodes in the network. Shortening the inter-block generation time brings higher block propagation delay, which increases the temporary forks rate ~\cite{temp}\cite{BlockPro}, which wastes the miner resource and makes the transactions wait longer. Alternatively, increasing the inter-transaction generation also increases propagation delay, affecting the confirmation waiting time of the transaction~\cite{prop}. Likewise, the peer formation strategies also impact the reachability of the transactions and blocks~\cite{simblock}\cite{BitINV}. The nodes forward new updates to the peer nodes, in which the number of peer nodes and the delay in between impact the amount of time needed to forward a message. The network elements delay, processing delay, and peer formation strategies affect the block's number of minutes to reach the maximum number of nodes. 

This paper aim to investigate the impact of these network-related parameters on the overall performance of the technology. However, it is difficult because the nodes are independent and anonymous, making it challenging to collect measurement data from the unknown nodes. For this reason, a testbed has been prepared to perform a measurement-based study. As a highlight, the testbed includes 104 raspbeery pis, six switches, two-blade racks. Each blade rack can hold up to 40 raspberry pis. Each raspberry device has Bitcoin Core 0.21.0 installation with additional scripts to automate transactions and block generation events. Through this testbed, a dataset has been gathered containing primary information about the chain, i.e., the ledger, and information that is not available from the ledger but measured from the local mining pool (mempool). Based on the collected dataset, an explorative study on the transaction characteristics of Bitcoin has been conducted.

The rest of the paper is organized as follows. The current state of the art is covered in Section~\ref{sec-stateArt}. Next, Section~\ref{sec-setup} illustrates the testebd setup and what kind of parameters considered. Then, Section~\ref{sec-workflow} illustrates the workflow of transaction handling in Bitcoin. Following that, how P2P topology formation and the strategies proposed are discussed in Section~\ref{topo}. Next, Section~\ref{sec-trans} and \ref{sec-trasconf} reports results gained from the analysis. Following that, Section~\ref{sec-fork} presents the impact of fork occurrence over the transaction confirmation time. Section~\ref{sec-dis} opens up a discussion on what has been observed in the analysis. Finally, Section VII concludes the paper.

\section{Related Work} \label{sec-stateArt}
There are several works related to studying the impact of bitcoin peer-to-peer (P2P) on the security and performance of the technology. Jean-Philippe et al.~\cite{p2pdiscovery} analyzed examining the resilience of bitcoin networks from churn, detection of Sybil nodes, dynamicity, and popularity of peers. Based on one month of observation, the study showed little churn in the network, no Sybil attack, and recent updates on tackling these issues had become effective. Taotao et al.~\cite{Etherp2p} developed an Ethereum network analyzer, Ethna, to analyze the P2P network. The analysis showed that the average degree of an Etherium node is 47, and the P2P network of blockchain such as Bitcoin degree of distribution follows a power-law. The network has the characteristics of a scale-free network.  

Muntadher et al.~\cite{LocalApproch} proposed locality-based approaches to improve the propagation delay on the P2P network.  This study considered clustering nodes in the exact geographical location, where the distance between is used as key on choosing which nodes to add as a peer. They showed that providing a less distance threshold would improve the transaction propagation delay with a high proportion. However, clustering with known deterministic distance may reduce the security of the network. Maryam et al.~\cite{p2ptoplogy} proposed a Bitcoin P2P topology discovery framework that tracks the information exchange to discover network topologies.  Based on 45 days' observation, the node distribution between the USA and China matches closely, while other parts of the world have fewer active public nodes to discover.  

Yahya et al.~\cite{BlockPro} proposed an analytical model to study the network delay and traffic delay in Bitcoin. The study considered the effect of the default number of connections and the block size on the performance of the Bitcoin network. Varun et al.~\cite{BTCmap} developed a fast and efficient framework named BTCmap to discover and map the Bitcoin network topology. The analysis indicates that the online peers' list remains valid (less than 1\% of changes) at 56 minutes 40 seconds. Otsuki et al.~\cite{RelayNetwork} showed that a relay network improves the propagation time of a block. In addition, the work showed that relay network decrease in the orphan block rate and the 50th percentile of block propagation time. However, the relay network's improvement of the orphan block rate became smaller as the Internet speed increased. Regarding the mining success rate,  it was demonstrated that the relay network did not significantly influence: the differences between utilizing and non-utilizing nodes were below 0.1 at any utilization rate. 

Most of the research work outlined above analyzes the discovery of Bitcoin P2P topology or develops a framework to crawl to the live Bitcoin network to statistics to discover the structure and security bridges of the technology. However, little has been investigated about the impact of peer selection, topology formation, and end-to-end delay on the transaction characteristics of Bitcoin. To this aim, we developed a testbed that mimics the real Bitcoin network, enabling us to experiment with collected data and make further analyses.

\section{Measurement setup and Node configuration}\label{sec-setup}
\begin{figure}[ht!]
\centering
  \includegraphics[width=0.8\linewidth, height=0.76\linewidth]{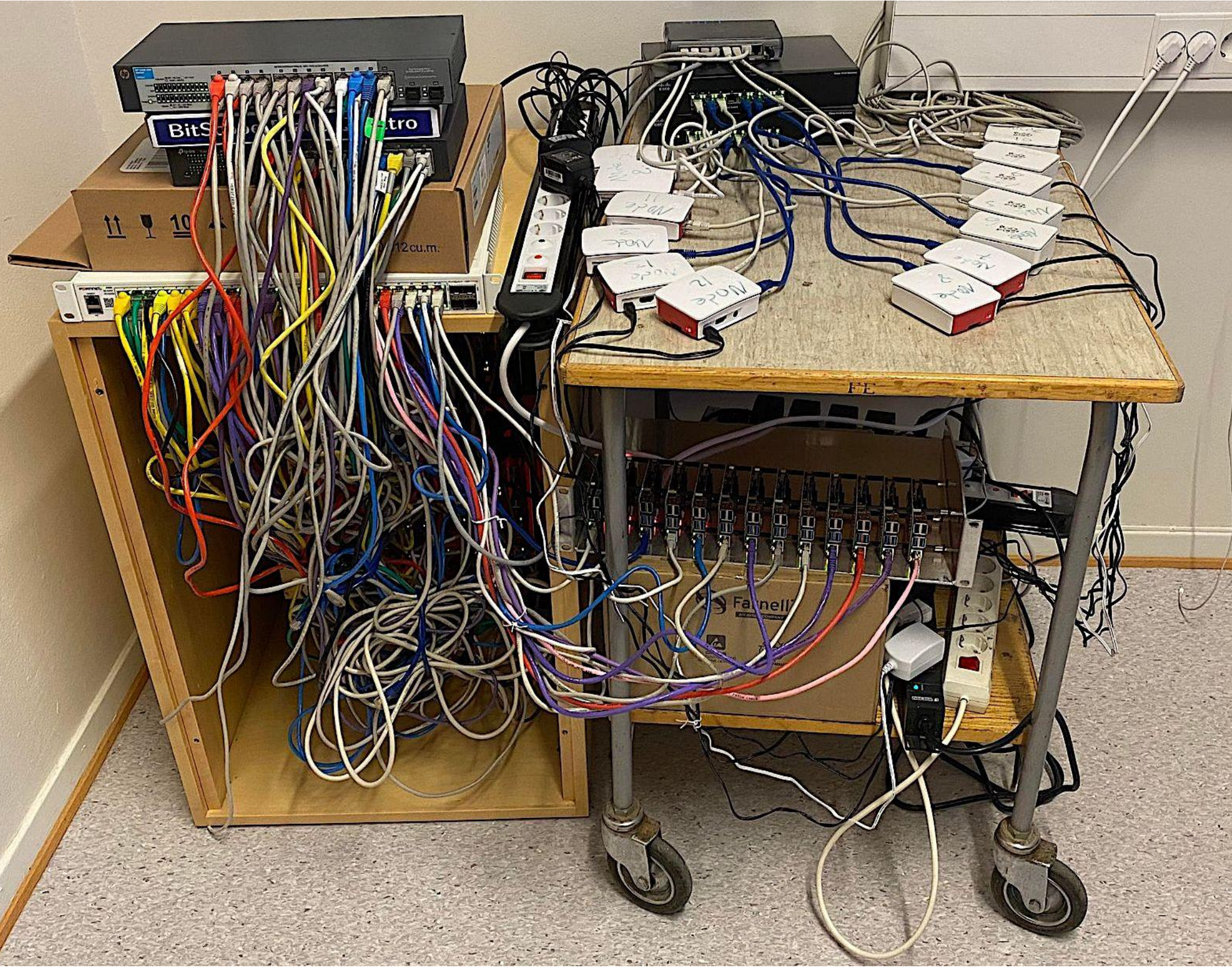}
   \caption{Testbed setup (104 Raspberry pis, 6 switches, 2 blade rack ) }
  \label{setup}
\end{figure}

For the measurement study, as shown in Fig. \ref{setup}, a testbed has been implemented to record information about Bitcoin transactions. The testbed includes 104 Raspberry pis, six switches, two-blade rack (each holding 40 Raspberry pis). Each Raspberry pis has an installation of a full Bitcoin core. 
\subsection{Node Configuration}
Raspberry Pi's are used as a full node that participates in addition, validations, and generating valid logs. These devices boot from an SD card. The SD card has Ubuntu Server Version 20.10 for the ARM architecture. In addition, the SD cards contain the scripts necessary to run the setup, for instance, scripts to start bitcoin daemon, adding topology and delay and generate transactions and blocks. 
\subsubsection{Network configuration}
Each node interface is configured with an IP address 192.168.xx.1/24. Subnetting with /24 may not be necessary to have a single node, but we plan to increase nodes per subnet for the future use case. Assigning such an IP address also mimics an actual Bitcoin node with its public address. Since each node becomes part of its network, we used VPP (Version 21.6 ) to perform routing between the nodes. It is an open-source software that provides high-performance switching and routing features for commodity hardware~\cite{VPP}. 

\begin{figure}[h!]
\centering
  \includegraphics[width=0.8\linewidth, height=0.5\linewidth]{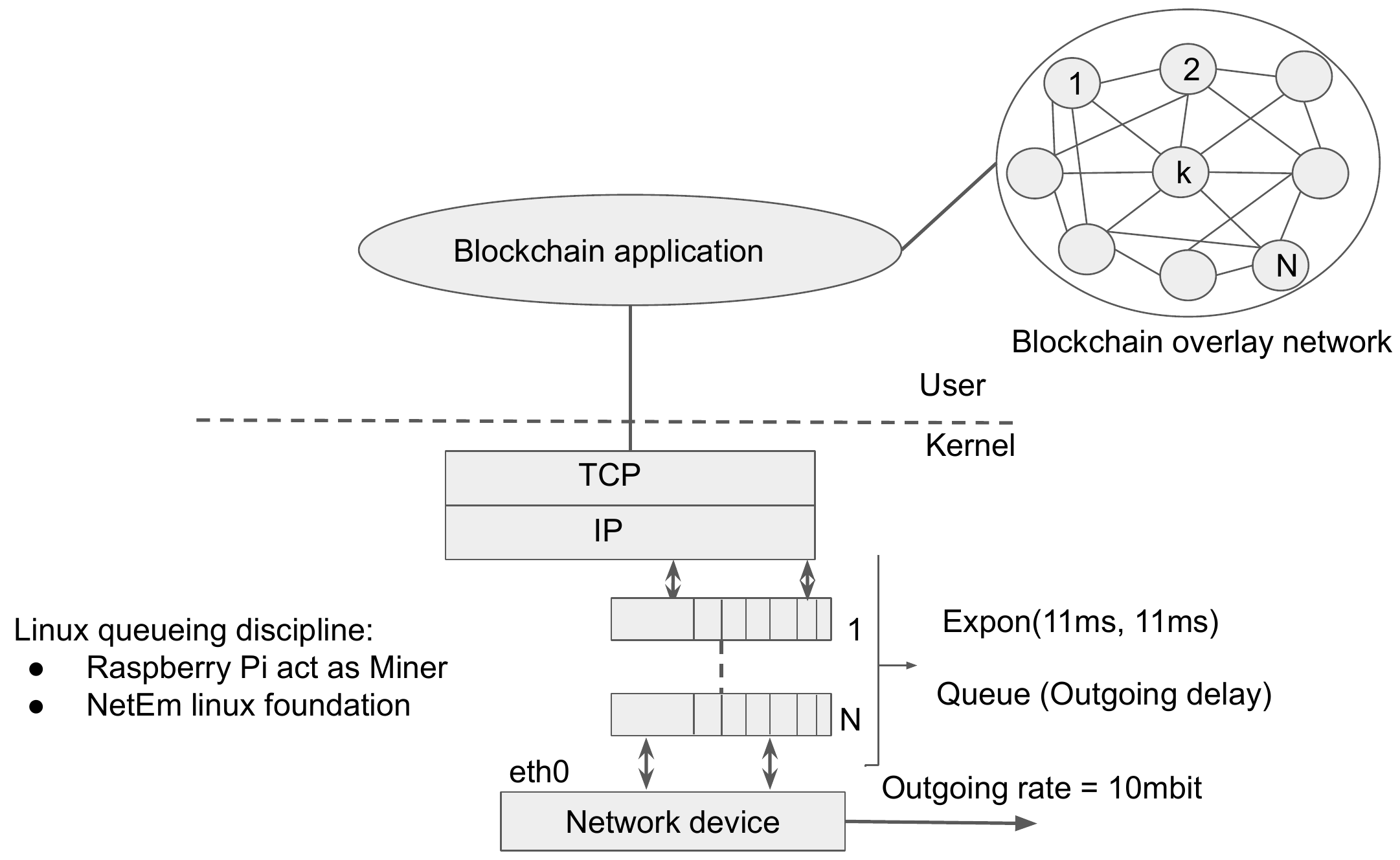}
  \caption{ Bitcoin node configuration} 
  \label{nodeNode}
\end{figure}

The basic architecture of Linux queuing disciplines is shown in Fig.~\ref{nodeNode}. The queuing disciplines exist between the protocol output and the network device, and the default queuing discipline is a simple packet FIFO queue. A queuing discipline is a simple object with two key interfaces. One queues packets to be sent, and the other released packets to the network device. The queuing discipline makes the policy decision of which packets to send based on the current settings. 
As Fig.~\ref{nodeNode} shows, the packet leaving each node adds delay to each packet that follows an Exponential distribution.  Since each node has an N peer list, we can also see N queues.  In addition, the bandwidth is limited with 10 Mbps capacity.  These configurations mimic the real Bitcoin network's peer list, and delay arises from the node and network capacity limitations.  To simulate a network of the whole Bitcoin network, we used NetEM. It provides Network Emulation functionality for testing protocols by emulating the properties of wide-area networks~\cite{netemE}. 


NetEm emulates the actual network traffic since the traffic characteristics are unpredictable. This emulator provides Normal and Pareto distributions~\cite{netemE, netem}. However, we believe the inter-packet delay follows an exponential distribution in our setup. This is another challenge since the NetEm does not provide this distribution but allows users to add their distribution. There are different ways to prepare a user-defined distribution. For instance, extracting the RTT values from ping statistics gives the mean and standard deviation, then using it in the NetEm command when activating the distribution table produced. This is easy to do between a few nodes. Our setup mimics the actual Bitcoin network of 5670-7279 active full nodes~\cite{p2pnetwork}\cite{p2ptoplogy}. The Bitcoin documentation states that a node chooses a peer within shorter latency. We generated random variates by inverse transform sampling of exponential distribution based on this fact and then used iproute2 marketable to create an exponential distribution. We set the delay (d) between 11 ms, and it is a shorter end-to-end delay to add nodes. This 11 ms is extracted from an independent full Bitcoin node~\cite{transHandling}, we calculated the delay between the eight peers from this node and took the minimum delay between the node, and its peer was 11 ms.  
\subsubsection{Node to node delay}
\begin{figure}[ht!]
\centering
  \includegraphics[width=0.75\linewidth, height=0.2\linewidth]{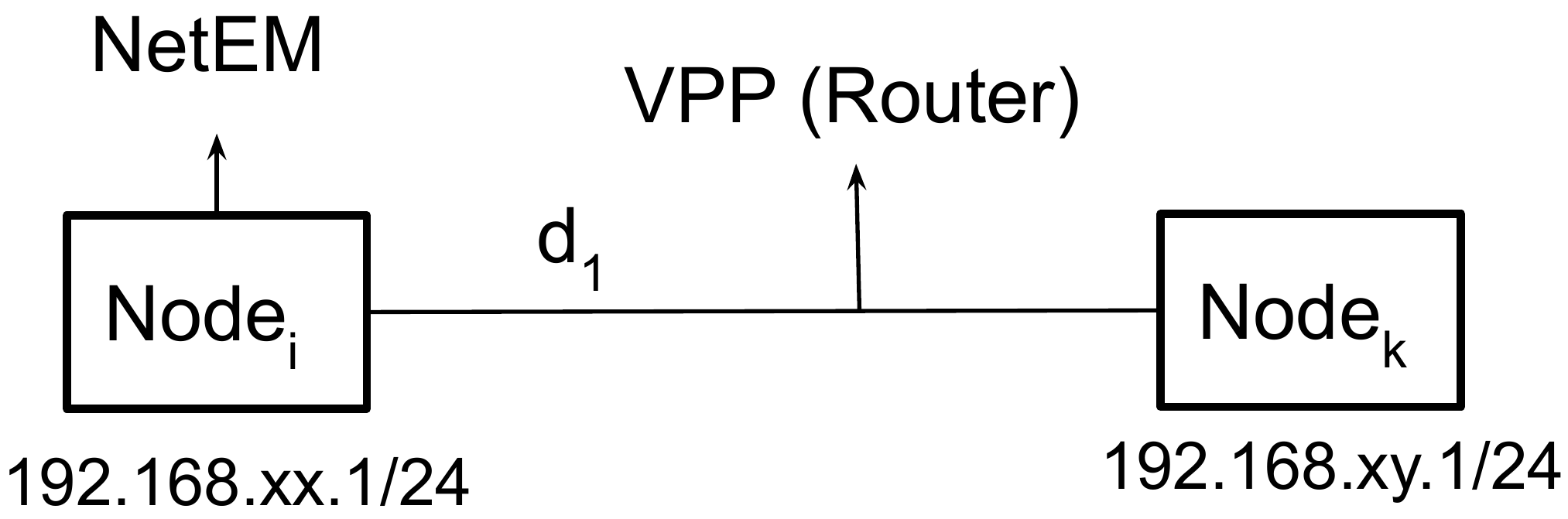}
  \caption{ Node adding delay  }
  \label{nodeToNode}
\end{figure}

In the previous subsection, we provided why NetEM is used to add delay and bandwidth limitation to emulate Wide Area Network (WAN). This section shows how independent nodes communicate with each other through an open-source software router Vector Packet Processing (VPP). Nodes add delay d to each outgoing packet. The outgoing packet passes through the router and reaches the destination. 
Fig. \ref{nodeToNode} demonstrates the node to node communication delay between Node$_i$ and Node$_k$ while the Dell computer is used as a router. The VPP open source software router is configured in Dell OPTIPLEX 9020, with a specification of Intel® 4th generation Core™ i7/i5 Quad Core, Ubuntu 20.04, 32GB memory, Integrated Intel® I217LM Ethernet LAN 10/100/1000, and 256GB storage capacity.
\subsection{Time synchronization}
\begin{figure}[ht!]
\centering
  \includegraphics[width=0.8\linewidth, height=0.45\linewidth]{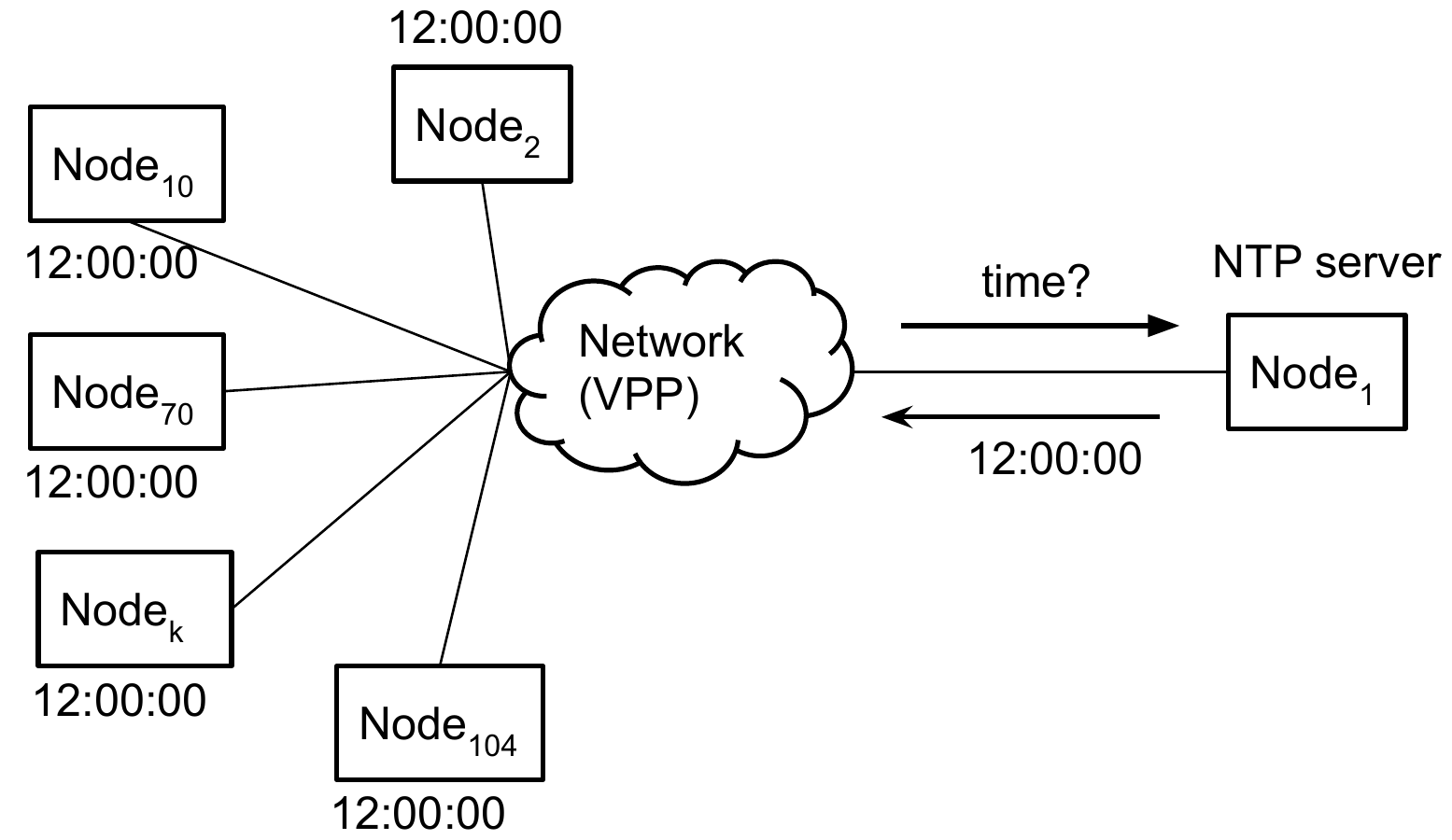}
  \caption{Time synchronization  }
  \label{NTP}
\end{figure}

The devices have to be time-synchronized to enable accurate time stamping by each node in the network.  For this reason, we used a well-known time synchronization application called Network Time Protocol (NTP). NTP is an application that allows computers to coordinate their system time~\cite{ntp2, ntp}. The implementation is in userspace rather than in kernel mode; however, its performance is much better than the other network time protocols~\cite{ntp}. Usually, it is available for most Linux distributions, which makes it easier to integrate with applications. We have 104 nodes that generate events that require accurate timing and synchronization. To this aim, we used NTP in our setup, node 1 acts as an NTP server, while the rest 103 nodes act as a client. The nodes' synchronize time means to set them to agree at a particular epoch with respect to coordinated universal time (UTC)~\cite{ntp2}. Fig.~\ref{NTP} shows how NTP is added to the setup. As we can see from the figure, node 1 is the NTP server, while the rest 103 nodes are the NTP clients.  

\subsection{Raspberry pi specification}
The Raspberry Pies devices are running the bitcoin protocol through Bitcoin Core 0.21.0. To identify them, each of them was given a unique number from 1 to 104. These devices act as full nodes, and a single device will be referred to as \textit{node n} where n is the given number. As we see from Table~\ref{tab:pi_models_specs}, in total, the setup has 93 Raspberry Pi 3 and 11 Raspberry Pi 4. There are some differences between Raspberry Pi 3 and Pi4 that are relevant for the setup. Raspberry Pi 4 Plus has a CPU clock speed of 1.5 GHz, 0.1 GHz more than Raspberry Pi 3, which has a clock speed of 1.4 GHz. Additionally, Raspberry Pi 3 has an Ethernet port with a maximum throughput of 300 {mbps} while Raspberry Pi 4 has Gigabit Ethernet. 

\begin{table}[ht]
\centering
\caption{Raspberry Pi Models}
\begin{tabular}{|l|l|l|}
\hline
 & Raspberry Pi 3 Model B+& Raspberry Pi 4 \\\hline
 
Processor & 1.4 GHz  & 1.5 GHz, 64 bit CPU\\ \hline

Memory & 1GB RAM & 1-4GB RAM     \\ \hline

WiFi& 2.4GHz Wireless LAN&2.4Ghz and 5Ghz Wireless \\ \hline

Ethernet & Gigabit Ethernet &Gigabit Ethernet \\\hline
SD card & 8-16 GB & 8-16 GB \\\hline
\# nodes & 93  & 11 \\\hline
\end{tabular}

\label{tab:pi_models_specs}
\end{table}


\section{Work flow of Bitcoin} \label{sec-workflow}
This section gives essential background on how Bitcoin handles transactions. In addition, how the nodes communicate and discover each other. 

\subsubsection{Workflow}
\begin{figure}[ht!]
\centering
  \includegraphics[width=\linewidth, height=0.35\linewidth]{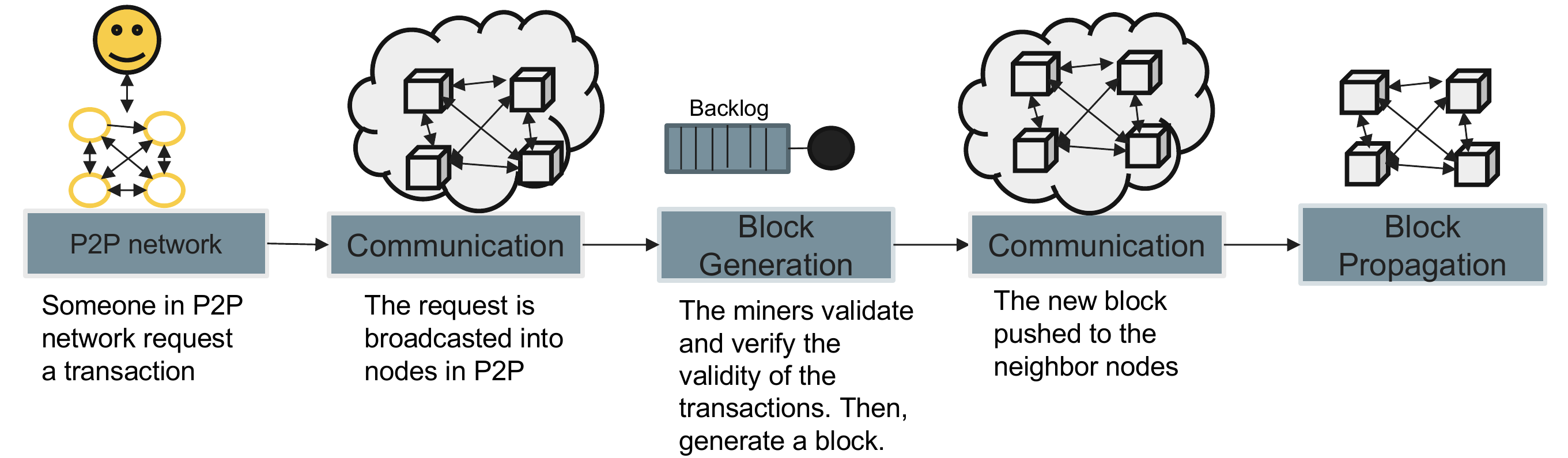}
  \caption{ Blockchain process flow }
  \label{process}
\end{figure}
Fig.\ref{process} illustrates the workflow of transaction arrival, block formation, propagation, and validation in Blockchain. Briefly, after transactions are generated by the users, they are sent to all full node validation nodes. Upon the arrival of a transaction at a full node, the node stores the transaction in its backlog (memory pool), waiting for confirmation. Besides, the node may choose unconfirmed transactions in the backlog to pack into a new transaction block. If the puzzle finding is successful, this newly generated block is added to the Blockchain. This information is sent to all the nodes. At each node, the validity of the newly generated block is checked. If the validity is confirmed with consensus, the updated Blockchain is accepted, and the new block transactions are validated. Such validated transactions are removed from the mempool at each full node that then repeats the above process.

\subsubsection{Node to node interaction}

\begin{figure}[ht!]
\centering
  \includegraphics[width=0.7\linewidth, height=0.43\linewidth]{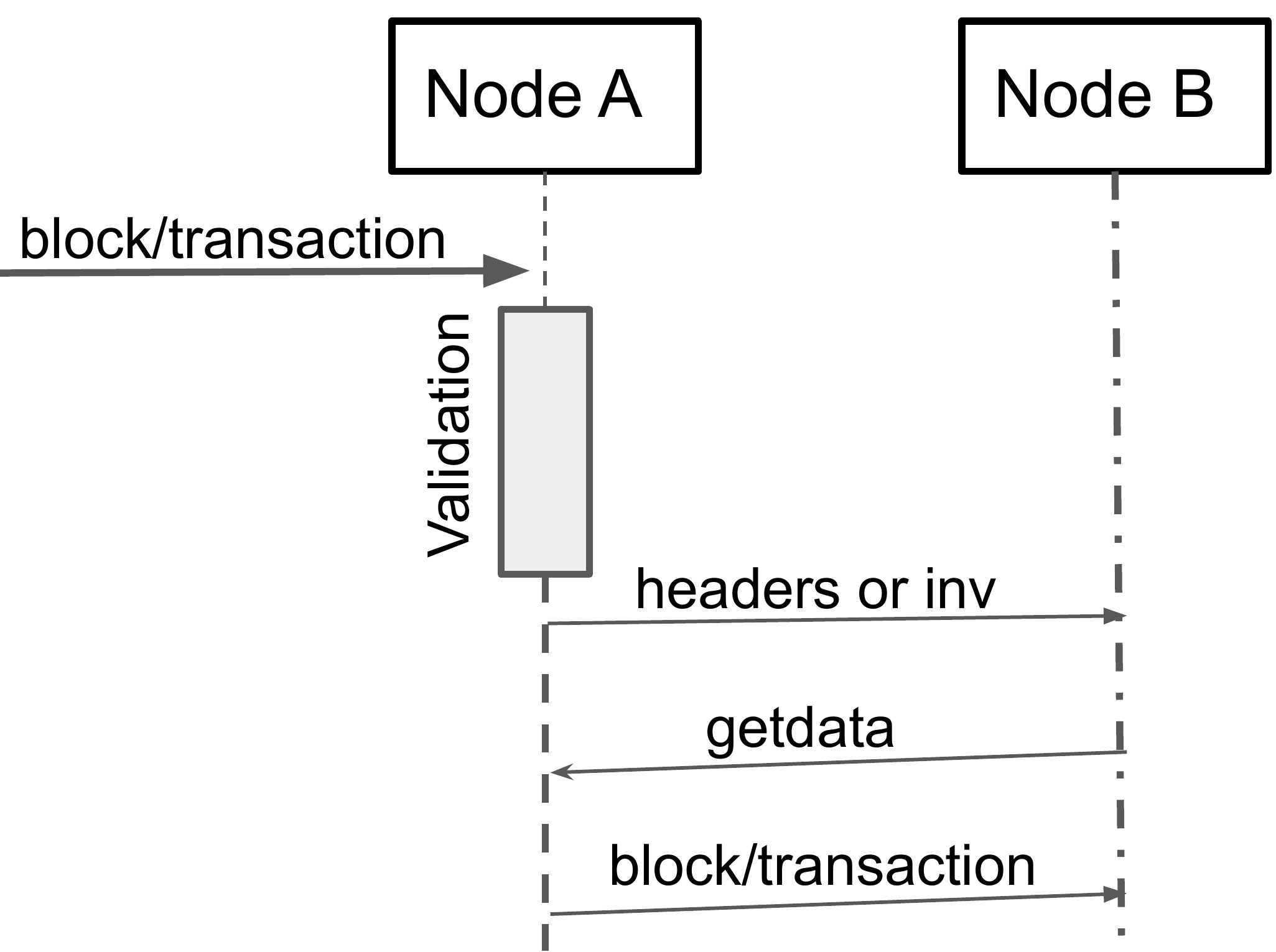}
  \caption{Legacy Relaying} 
  \label{nodeLegacy}
\end{figure}

Bitcoin nodes form a peer-to-peer network, while each node by default can have eight peer list. It is a logical link that allows peers to push/pull new updates to the neighbors. Fig.~\ref{nodeLegacy} shows node to node message exchange sequence. The new arrival block or transaction picked up node A. Then,  a block/transaction is validated (the grey bar) by Node A, who then sends an inv message to Node B requesting permission to send the block. Node B replies with a request (getdata) for the block/transaction, and Node A sends it.




\subsubsection{Network Discovery}
A Bitcoin node is allowed to maintain up to 132 connections (maxconnections) as default, of which 8 are outgoing connections and rest are incoming connections. Peers listen on port 8333 for inbound connections. When node wants to join the network, as it is a public blockchain, and node one uses DNS names (called DNS seeds) hardcoded into Bitcoin Core. From this point, the new node updates its peer list by discovering nodes close by In this way, new nodes select peers that are part of the network. This peer formation is called distance-based since it highly depends on adding nearby nodes. This peer list uses as a reference list to send an inventory or receive messages from the neighbor nodes. After the node joins the network, it can take part in propagation, consensus, and block generation. These nodes act as a full node, which means the users/owner can create a transaction and create a block, forward the new updates to the network. Each block created by the nodes that are valid enough to be included in the chain will contain the hash of previous records of the blocks. Blocks that are created but ignored by the network become orphan blocks. Mostly these blocks become fragments that will never be used but waste all the computation cost and resources. 

\subsubsection{Peer list}
Nodes can have up to 132 connection lists. This is the combination of incoming and outgoing peers. When a node initiates the connection, it is called outgoing, or if the connection initialization comes from other nodes, it is incoming bound. The number of peers (P) represents the number of outgoing peers of each node. The total connection list is the sum of P outgoing peers plus incoming peers (Q). In this work, the peer list ($p_l$) is 2P.  
\section{Setup input}
This section describes the input parameters such as inter-transaction generation time, inter-block generation time, and node to node delay added to the network.

\begin{figure*}[h!]
    \centering
    
    \subfigure[Distance-based approach ]
    {
    \includegraphics[width=0.45\linewidth,height=0.37\linewidth]{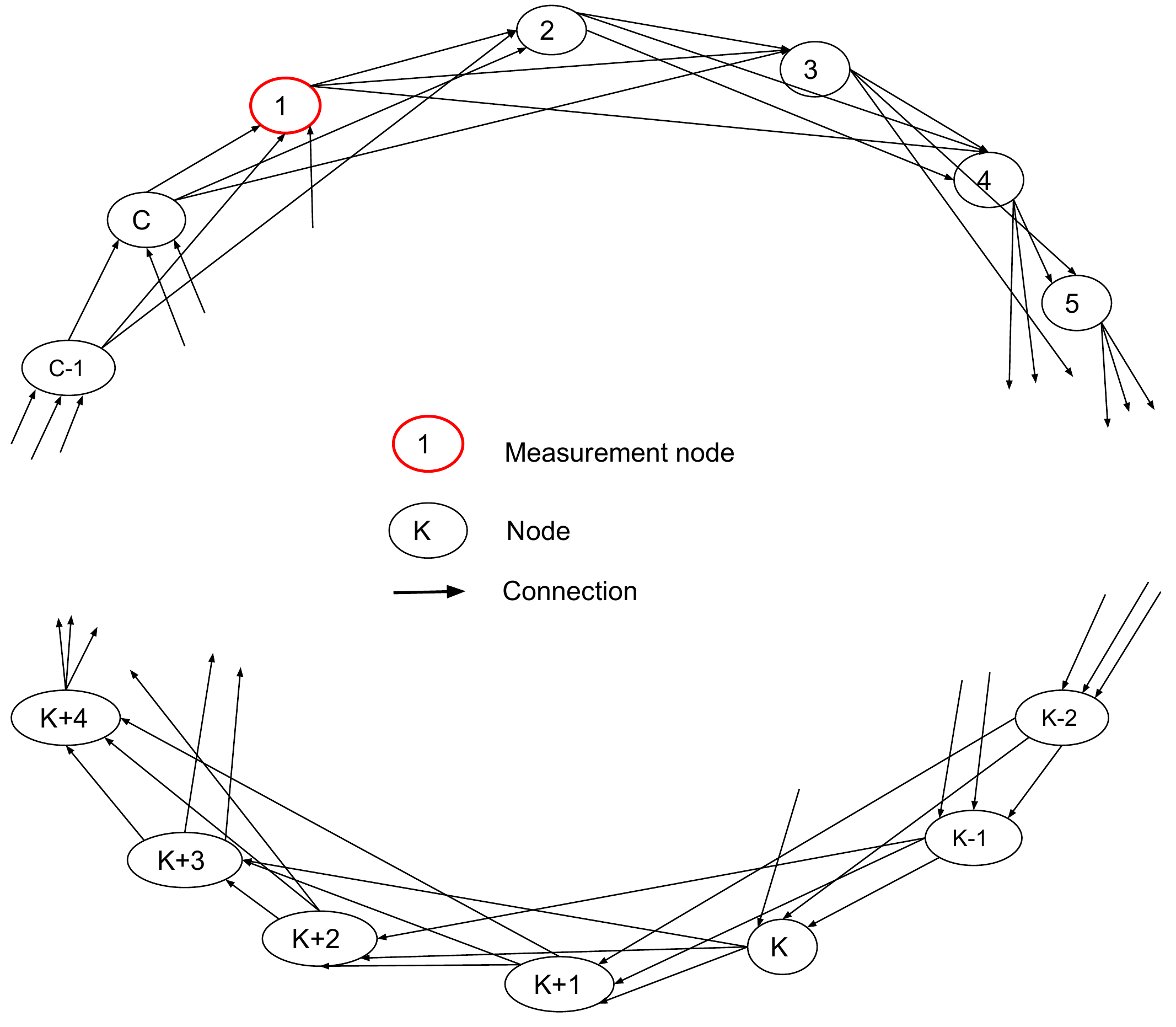}
   \label{distance}
    }
    \subfigure[Mixed-based approach ]
    {
    \includegraphics[width=0.45\linewidth, height=0.37\linewidth]{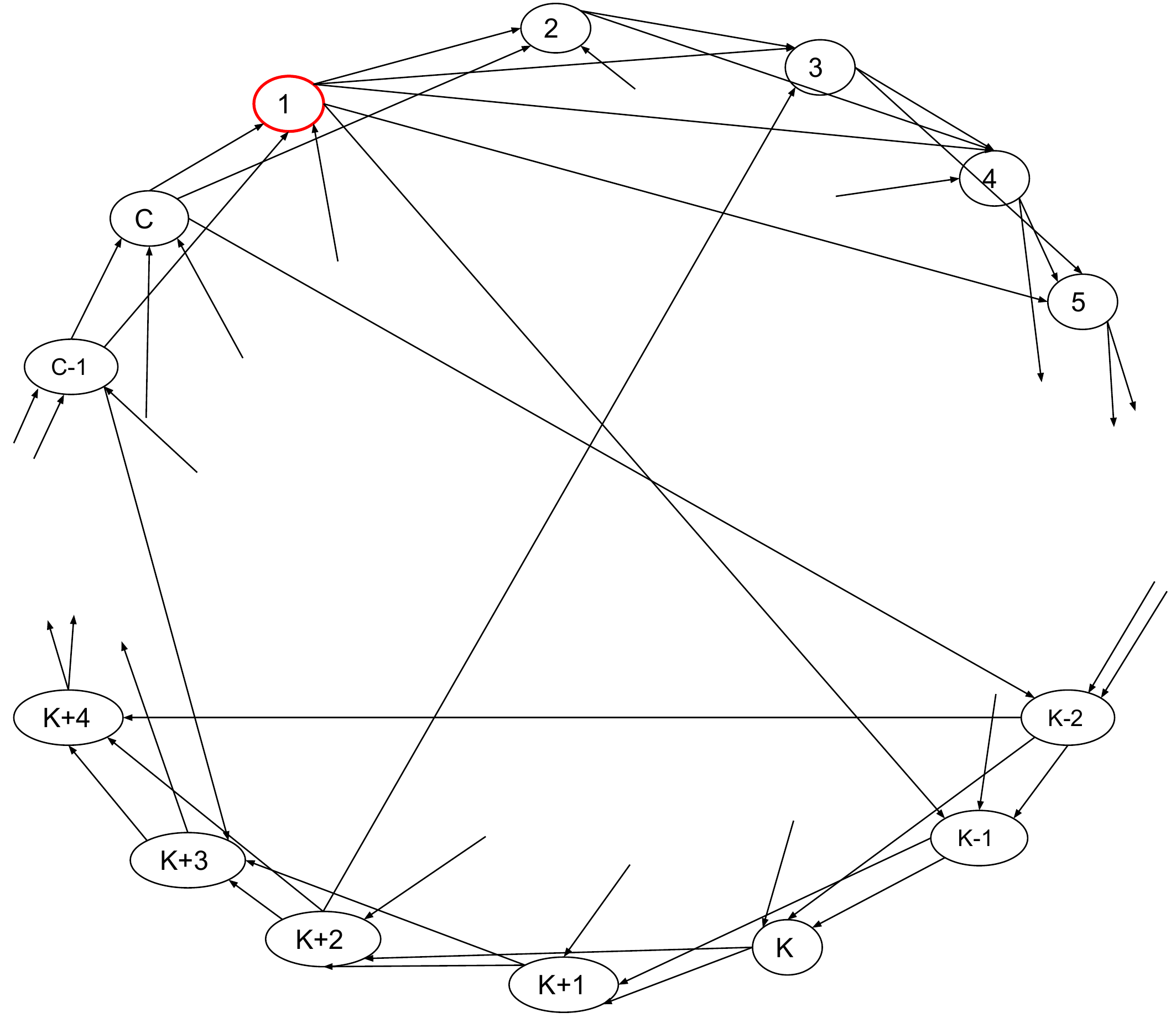}
   \label{mixed}
    }
   \label{nodeComm}
  \caption{Bitcoin overlay network example (P=3), while considering only outgoing links }  
\end{figure*}


The transaction and block generation events must also include similar characteristics to mimic the entire Bitcoin network. The transactions inter-arrival to the measurement node follow an exponential distribution~\cite{trasactionConfirmation}\cite{transBit}. Similarly, the inter-block generation also follows an exponential distribution~\cite{BitcoinBlockchainDynamics}\cite{transBit}. 

\subsubsection{Transaction inter-generation time}
\begin{algorithm}
\small
\caption{Generate transaction}
\begin{algorithmic}[1]
\Procedure{Poisson}{$\lambda(t),T_d$}       
    \State Initialisation: $T_t= timenow() + T_d$ 
    \State Condition: $T_d \leq T_t$
    \While{$True$}  
        \State $w_t$ $\sim$ negExp($\lambda(t)$) 
        \If{$timenow() + w_t < T_t$} 
        \State $time.sleep(w_t)$
        \State $generateTransaction()$ 
        \EndIf
    \EndWhile  
\EndProcedure
\end{algorithmic}
\label{AlgoTrans}
\end{algorithm}

Each node acts as a full Bitcoin node that creates, validates, and propagates transactions and blocks. To this aim, nodes have a script that generates transactions and blocks following an exponential distribution. The script accepts duration and the inter-generation interval in terms of seconds as an input parameter, as illustrated in algorithm~\ref{AlgoTrans}.  $\lambda(t)$ is the inter-generation time ($t_g$$_{i+1}$ - $t_g$$_i$) in seconds for each node. Furthermore, $T_d$ is the total duration of running time in seconds. The result of the inter-generation time distribution follows an exponential distribution. 


\subsubsection{Block inter-generation time}

Bitcoin network generates a block on average 10 minutes. This makes the recent block propagate to the network before the next generation. Bitcoin adjusts the difficulty after 2016 blocks are generated to control the average inter-block generation time. Although this is true for live Bitcoin nodes, the Bitcoin core regtest mode has difficulty close to zero, which means there is no difficulty generating a block. However, to mimic the real Bitcoin network, we developed a script that produces a block on average ten minutes. Overall, we have 104 nodes, which mean a block is generated in 103*600 second (61800), the remaining 1 node is measurement node. Similar to the previous transaction generation case, in here also the Algorithm~\ref{AlgoBlock} takes the inter-generation interval and duration of the simulation in seconds as an input. 

\begin{algorithm}
\small
\caption{Generate Block}
\begin{algorithmic}[1]
\Procedure{Poisson}{$\lambda(t),T_d$}       
    \State Initialisation: $T_t= timenow() + T_d$ 
    \State Condition: $T_d \leq T_t$
    \While{$True$}  
        \State $w_t$ $\sim$ negExp($\lambda(t)$) 
        \If{$timenow() + w_t < T_t$} 
        \State $time.sleep(w_t)$
        \State $generateBlock()$ 
        \EndIf
    \EndWhile  
\EndProcedure
\end{algorithmic}
\label{AlgoBlock}
\end{algorithm}

\subsubsection{Node-to-node delay}
In the actual Bitcoin network, nodes are distributed across the globe, which geographically and domain-wise isolated from each other. Since the underlying network infrastructure is providing the communication platform and the actual network traffic is unpredictable. It is common to consider a distribution that captures the network delay between two participating ends. To mimic the delay that arises from the network element and distance between the participating nodes. We introduced a delay (d) that follows an exponential distribution with the shorter mean of 11 ms. 


\section{Network topology}\label{topo}
To study the topological impact, we consider three peer formation cases: distance, random, and mixed. 

\subsection{Distance-based peer selection}
Distance-based peer selection approach enables peers to form closeby neighbor peer creating P2P topology. A Full Bitcoin node can have eight peers by default, but It can have 132 connection link points. The distance metric depends on adding nodes nearby. 

\begin{algorithm}
\caption{Distance-based}
\small
\begin{algorithmic}[1]
\Procedure{Distance}{$P, k, C$}       
    \State $p_l$= \{l|l= (k+i)\; mod\; C, i=1, \ldots,P\}
\EndProcedure
\end{algorithmic}
\label{AlgoDistance}
\end{algorithm}

Algorithm~\ref{AlgoDistance} illustrates the distance-based peer selection method. The procedure takes the number of peers to add (P), the current node (k), the total number of nodes (C). The algorithm add peer that are closeby.   


\subsection{Random-based peer selection}
Unlike the first distance-based approach, the random-based method does not depend on the proximity of nodes, instead on the random selection of the peer to add.  Even-though, Bitcoin is a distributed P2P technology where each node acts and does as an independent node, in which it has less knowledge on the global distribution of the nodes. For the random-peer selection method, we consider nodes know the number of Full active nodes in the network they are participating in. Similar to the distance-based approach, Algorithm~\ref{AlgoRandom} illustrates the random peer selection method. The procedure takes the number of peers to add (P), the current node (k), the total number of nodes (C). The method adds randomly selected nodes as its peer list. 


\begin{algorithm}
\small
\caption{Random-based}
\begin{algorithmic}[1]
\Procedure{Random}{$P, k, C$}       
    \State Initialisation: $p_l= \{\}, p_c= \{1, \ldots,C\}\setminus\{k\}$ 
    \For{$i=1$ step $1$ until $P$}
    \State $p_l\Large{\gets} p_l \cup ($RANDOM$(p_c\setminus p_l))$
     \EndFor
\EndProcedure
\end{algorithmic}
\label{AlgoRandom}
\end{algorithm}

\subsection{Distance + Random (Mixed)-based peer selection}
The third case is to combine distance-based and random-based approaches. In these combinations, the distance-based method adds n-1 peers and the random-based approach adds the last node by choosing randomly. This is to introduce a random link to the distance-based peer list. Same as to the previous two approaches, Algorithm~\ref{AlgoMixed} illustrates the mixed peer selection method. The procedure takes the number of peers to add (P), the current node (k), the total number of nodes (C). As discussed in the previous subsection, the method adds the n-1 nodes based on a distance-based approach. The random-based approach adds the last node. 

\begin{algorithm}
\small
\caption{Mixed-approach}
\begin{algorithmic}[1]
\Procedure{Random}{$P, k, C$}       
    \State  $ p_l= \{l|l= (k+i)\; mod\; C, i=1, \ldots,P-1\}$
    \State  $p_c= \{1, \ldots,C\}\setminus\{k\} \setminus p_l$ 
    \State $p_l\gets p_l \cup $RANDOM$(p_c)$
    \EndProcedure
\end{algorithmic}
\label{AlgoMixed}
\end{algorithm}

From this point on forwarding, we use random to represent a random-based approach, normal for distance-based default approach, and mixed for the approach that mixes the two approaches. 

\section{Setup validation}
This section relates the timings in the testbed with those in  the live Bitcoin network. 


\begin{figure}[t!]
    \centering
    
    \subfigure[Bitcoin live full node ]
    {
    \includegraphics[width=0.85\linewidth,height=0.45\linewidth]{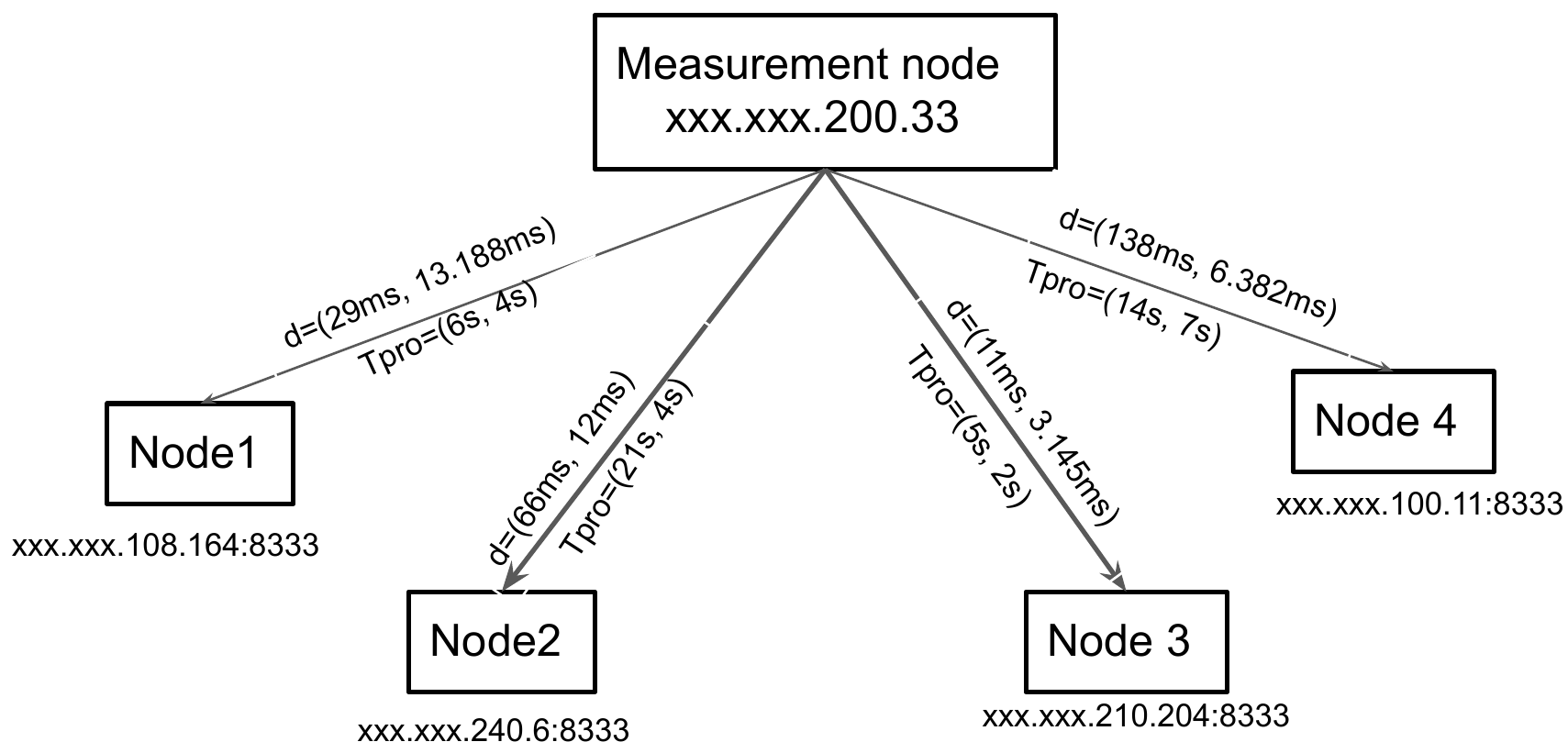}
   \label{TranProBitNode}
    }
    \subfigure[Testbed ]
    {
    \includegraphics[width=0.85\linewidth, height=0.45\linewidth]{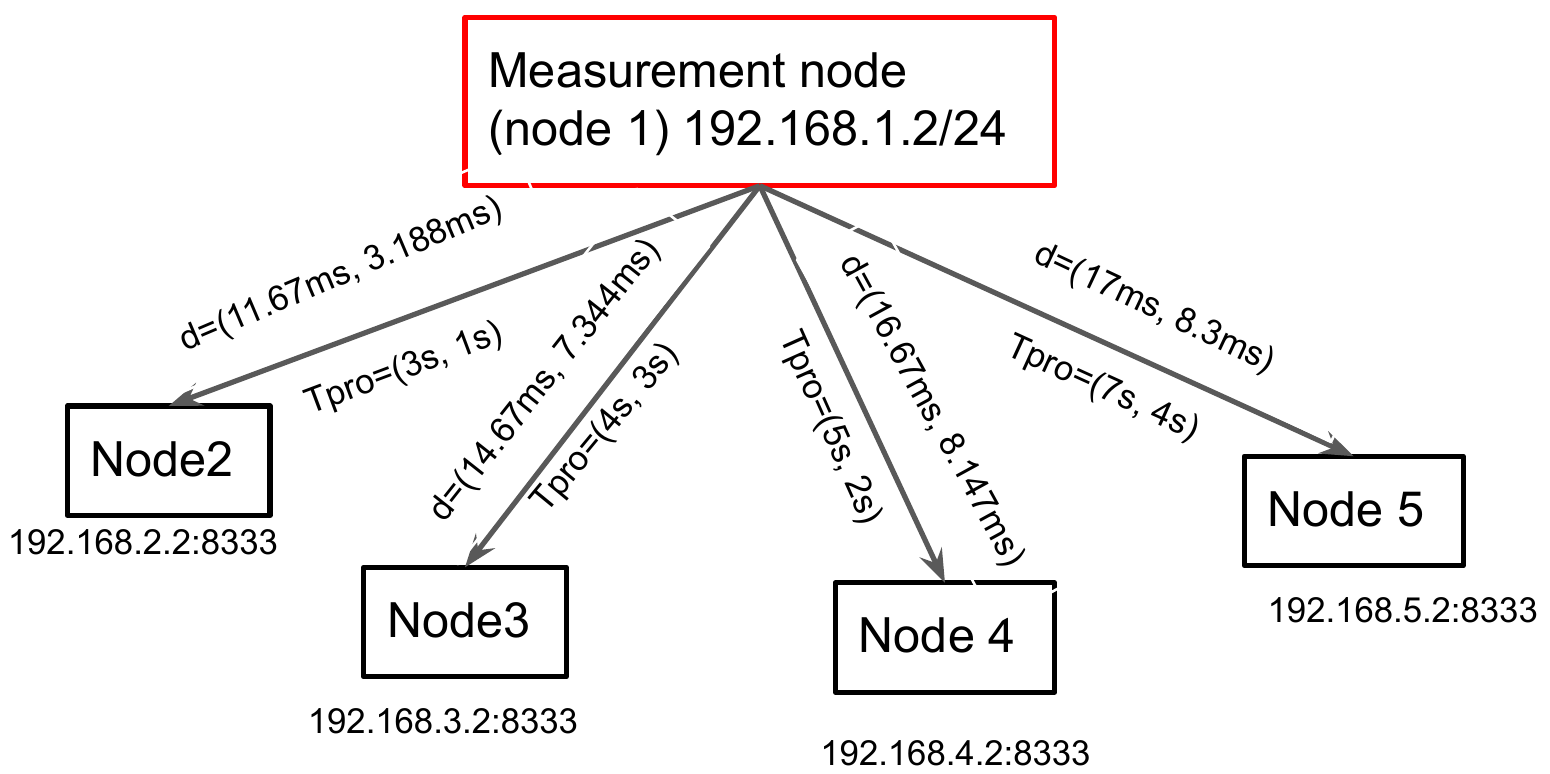}
   \label{TranProNodeTestbed}
    }
   \label{NodeTpro}
  \caption{Transaction propagation delay between active full nodes, where $\lambda = 3$  transactions per second per network, eight outgoing peers per node}  
\end{figure}

\subsection{Node to node delay}
In our previous work~\cite{transHandling}, an independent Bitcoin full node was deployed to collect transactions and block related feature sets. We used observations from this node to validate some of the input parameters and results. For instance, our nodes have 132 connected nodes. Eight of these nodes are peer nodes, while the rest are incoming bound nodes. The average ping delay between these nodes from the Bitcoin application is 156.20 ms with a variance of 152.23 ms. This ping is handled in a queue with other commands in the application layer to include the processing backlog. However, we also conducted further analysis to ping these nodes from outside of the Bitcoin core, resulting in an average of 80 ms second in deference. This 80 ms accounts for processing backlog. 

As mentioned in the previous section, the eight peers are more important than others. These peer nodes synchronize more often than the other 125 incoming bound nodes. For this purpose, we conducted an independent investigation to see the delay between our node to eight peer nodes. Our analysis shows that the minimum delay between our node and the other nodes is 11ms with a variance of 7ms. This 11 ms delay is used in our setup as a minimum delay guarantee between nodes.  
\subsection{Information propagation}
This subsection investigated how fast a transaction propagates in the Bitcoin network and how the number of nodes impacts this. We considered four publicly available nodes to collect mempool state and compare it with our node. Fig.~\ref{TranProBitNode} shows the delay between our node with four peer nodes that provide their state of the mempool. The figure shows only four out of eight nodes because the remaining four nodes were unreachable. As we can see from the figure, transaction propagation between nodes ranges from 13 to 20 seconds~\cite{transPro}. This is mainly because the P2P communication protocol makes processing check the validation of each transaction before forwarding an Inv message to its peers. At the same time, nodes that received the Inv message have to check if the transaction is at the mempool or seen before inside a block. The node sends a getdata message and gets the new transaction when the check is completed. Even though the delay between nodes is less than 100 ms, processing a transaction takes longer.

The analysis also shows that the number of transactions waiting in the mempool varies between peers in an instant of time. However, the difference between our node and the above four nodes used for the analysis ranges from 600-700 transactions. For instance, each node has 1566, 3976, 3000, 2244, and 2300 transactions waiting at the mempool. 

We tested out the testbed based on the live Bitcoin full node observation to see if similar transaction characteristics occurred. As we can see from Fig.~\ref{TranProNodeTestbed}, the transaction propagation delay between the measurement node, node 1, to its four peers is below 7s. This demonstrates that the timings in testbed are similar to those in the real Bitcoin network. 

\subsection{Inter-block generation and Inter-transaction arrival time}
The average inter-block generation time is close to 10 minutes in the actual Bitcoin network. After the 2016 blocks, the difficulty of solving the puzzles increases to make sure nodes generate on an average of 10 minutes so that the new block reaches the maximum number of nodes in the network. 
Similarly, the transaction inter-arrival time to the mempool also follows exponential distribution~\cite{transHandling}\cite{transBit}. These parameters are considered in our setup as an input parameters.

\section{Measurement data collection}
\begin{figure}[ht!]
\centering
  \includegraphics[width=0.9\linewidth, height=0.86\linewidth]{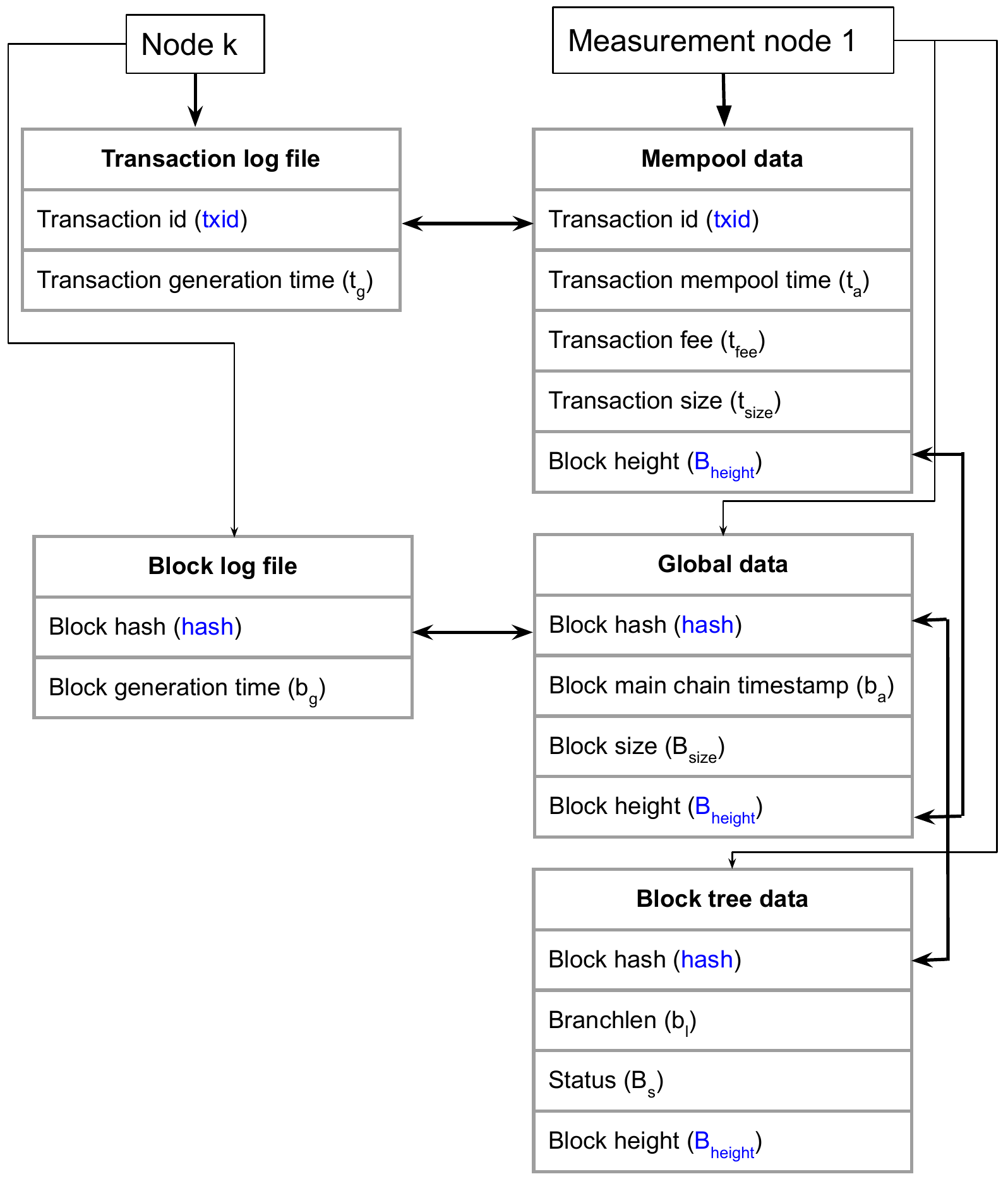}
  \caption{Data collection} 
  \label{datacollect}
\end{figure}

A dataset consisting of four parts has been collected by the testbed, as show in Fig.~\ref{datacollect}. One aspect of the dataset records each node transaction and block generation events. When a node Node k, where k $\in [2, 104]$, generates transactions, it records a log about the transaction generation time ($t_g$) and transactions id ($txid$). Similarly, when the node generates a block, it records the block generation time($b_g$) and block hash ($hash$). The second part of the data contains information about the transactions arrival time at mempool ($t_a$), transaction size ($t_{size}$), transaction fee ($t_{fee}$), transaction id ($txid$), and block height ($B_{height}$). The block height ($B_{height}$) in which the transactions belong can be empty or number depending on if the transaction is added to the block or just new arrival. The third part collects information about the block from the main chain, such as block hash ($hash$), block size ($b_{size}$), block time ($b_t$), and block height ($B_{height}$). The fourth part of the data collection contains extracted details about the block tree of the chain, such as Block hash ($hash$), Block height ($B_{height}$), Branchlen ($B_l$), and Status ($B_s$). The Branchlen is the length of the brach in the block tree. It holds 0 for the main chain or number, indicating the length of the soft fork in terms of the number of blocks in the side chain. The Status ($B_s$) indicates the Status of the block, whether it is active, part of the main chain, valid-fork means a block is a fork or invalid block, which means the block is not valid enough to be a candidate. 

The second, third and fourth part of the data is collected from a single node. This node is considered a measurement node, in our case,  node 1 is measurement node. Node 1 is part of the network invalidation and processing transaction at the mempool, but it does not generate transactions or blocks. Instead, it collects information about the transactions from its mempool (Mempool data). When the simulation times are over, it also extract information about valid blocks from the main chain (Global data, Block tree data). 

Fig.~\ref{datacollect} demonstrates the collected feature set from the nodes. As we can see from the figure, measurement node 1 collect information about the state of the mempool and keep track of the status of the main chain. It also illustrates the primary key used to link the data set from each device with node 1. 
By using the datasets, we performed analysis on transaction propagation ($t_a-t_g$) time and confirmation time ($b_{g(i+6)}$ - t$_g(x)$), where transaction $x$ goes into block $i$ and  $i+6$ represent when the transaction is six-block deep into the main chain.

In addition to the above-collected information, we also extracted the state of the block tree. This information includes which block is fork ($hash$), at which height this event happened ($B_{height}$), and the number of the block within the same branch ($B_{branchlen
}$). We used these extracted feature sets to count the number of forks that happened while considering different peer formation strategies and how they impact the confirmation time of transactions inside a fork block.  These datasets are downloaded and post-processed after the simulation period is completed.

\section{Transaction propagation time}\label{sec-trans}

This section reports results and observations from an exploratory analysis of the collected data.  This part covers the impact of transaction intensity while illustrating the impact of peer lists per node. 

\begin{figure}[ht!]
\centering
  \includegraphics[width=0.9\linewidth, height=0.36\linewidth]{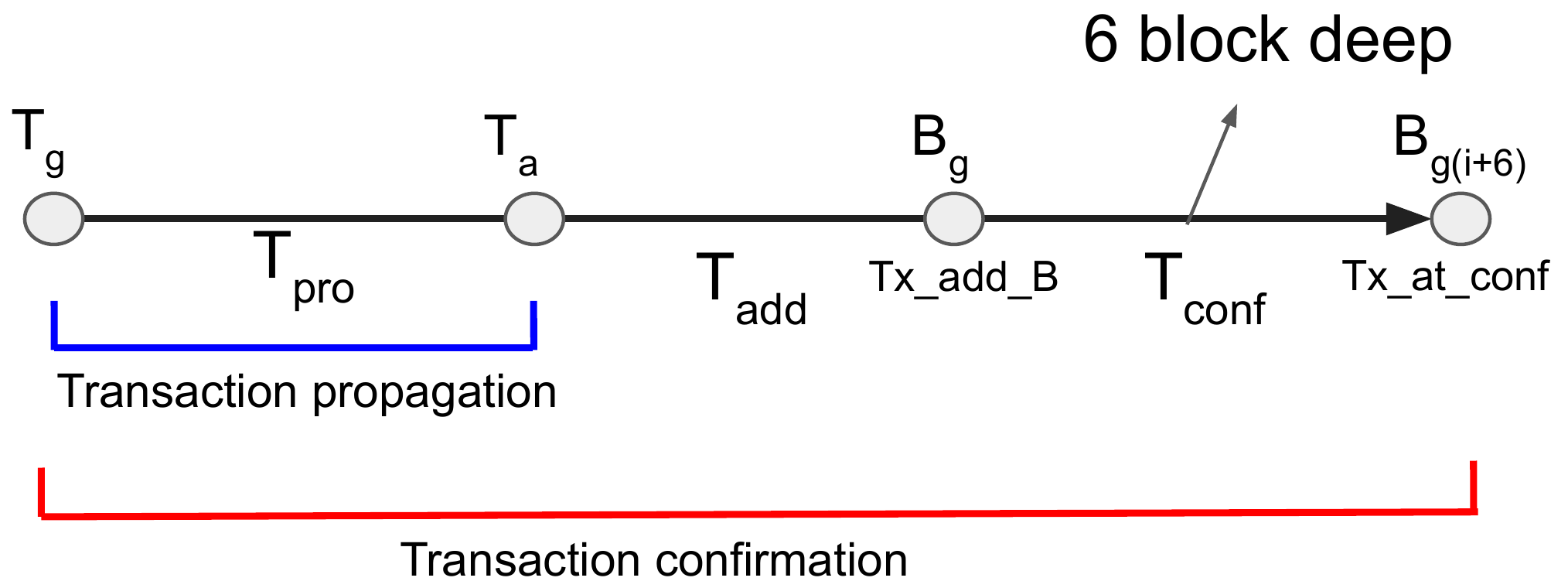}
   \caption{Transaction propagation and confirmation time sequence  }
  \label{TranProConfNodeseq}
\end{figure} 

Bitcoin uses a gossip-like protocol to broadcast updates throughout the network~\cite{p2ptoplogy}. When nodes receive new transactions, it validates and verifies the validity of the transactions, then sends an Inv message to peer nodes to notify them if the peer nodes want these new transactions. Then push the transaction to the peers. Due to this continuous process, a delay in transaction propagation happens. It combines validation time and the time it takes to disseminate the transaction. Fig.~\ref{TranProConfNodeseq} shows a time sequence of the life cycle of transactions. In this section, we focus on the transactions propagation, and this is the time transaction generated ($t_g$) until it reaches the memppol of measurement node, in our case, node 1. The time difference between $t_g$ and $t_a$ is the propagation time, where $t_a$ is the time transaction arrived at the memppol of node 1, and $t_g$ is the time of the transaction generated by one of the nodes (2-104). As we can see from Fig.~\ref{TranProConfNodeseq}, the red line indicates the time length of transaction propagation time.

\begin{figure}[ht!]
    \centering
    \subfigure[$P=8$]
    {
    \includegraphics[width=0.45\linewidth, height=0.4\linewidth]{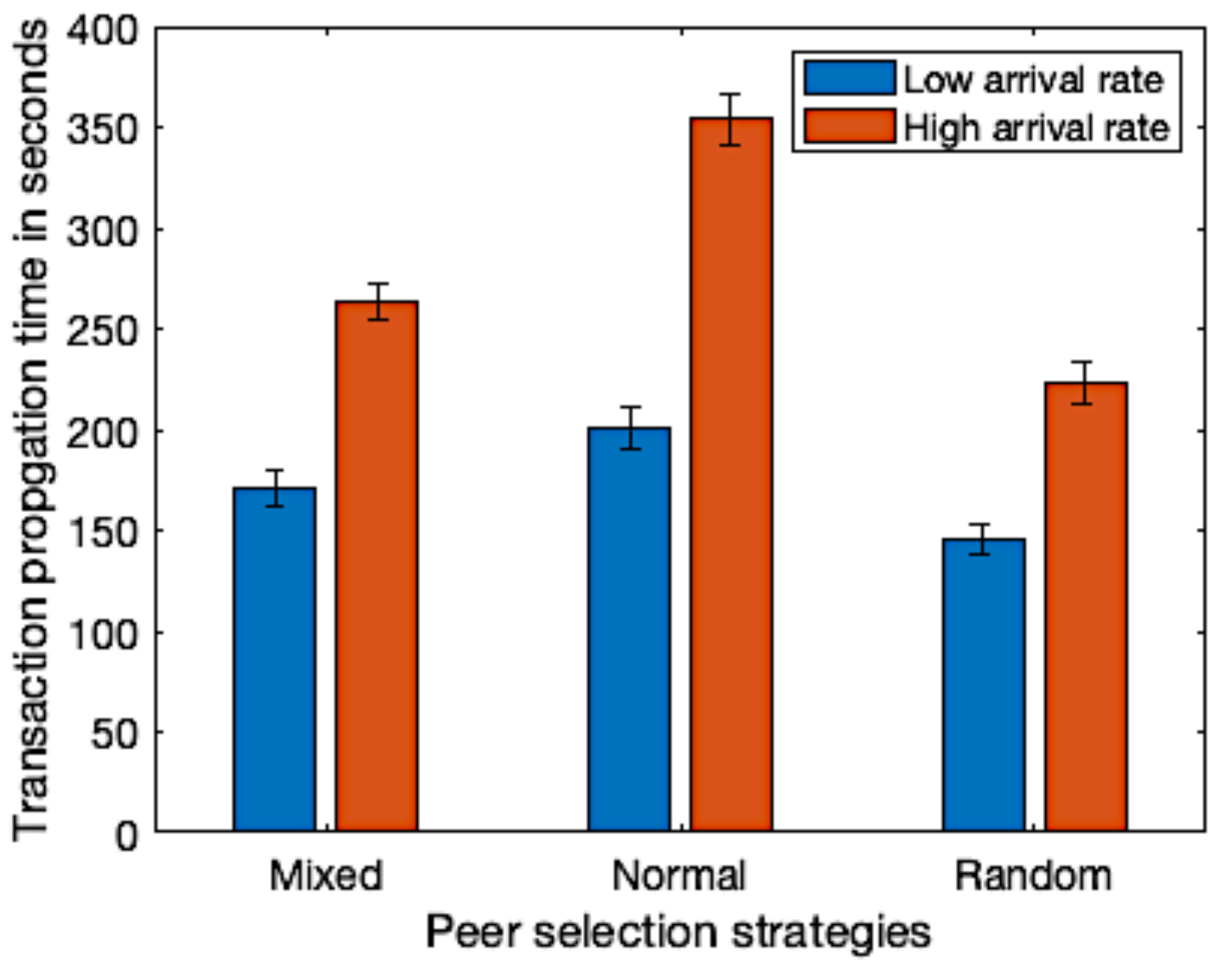}
   \label{TranProComp}
    }
    \subfigure[$P=4$] 
    {
    \includegraphics[width=0.45\linewidth, height=0.4\linewidth]{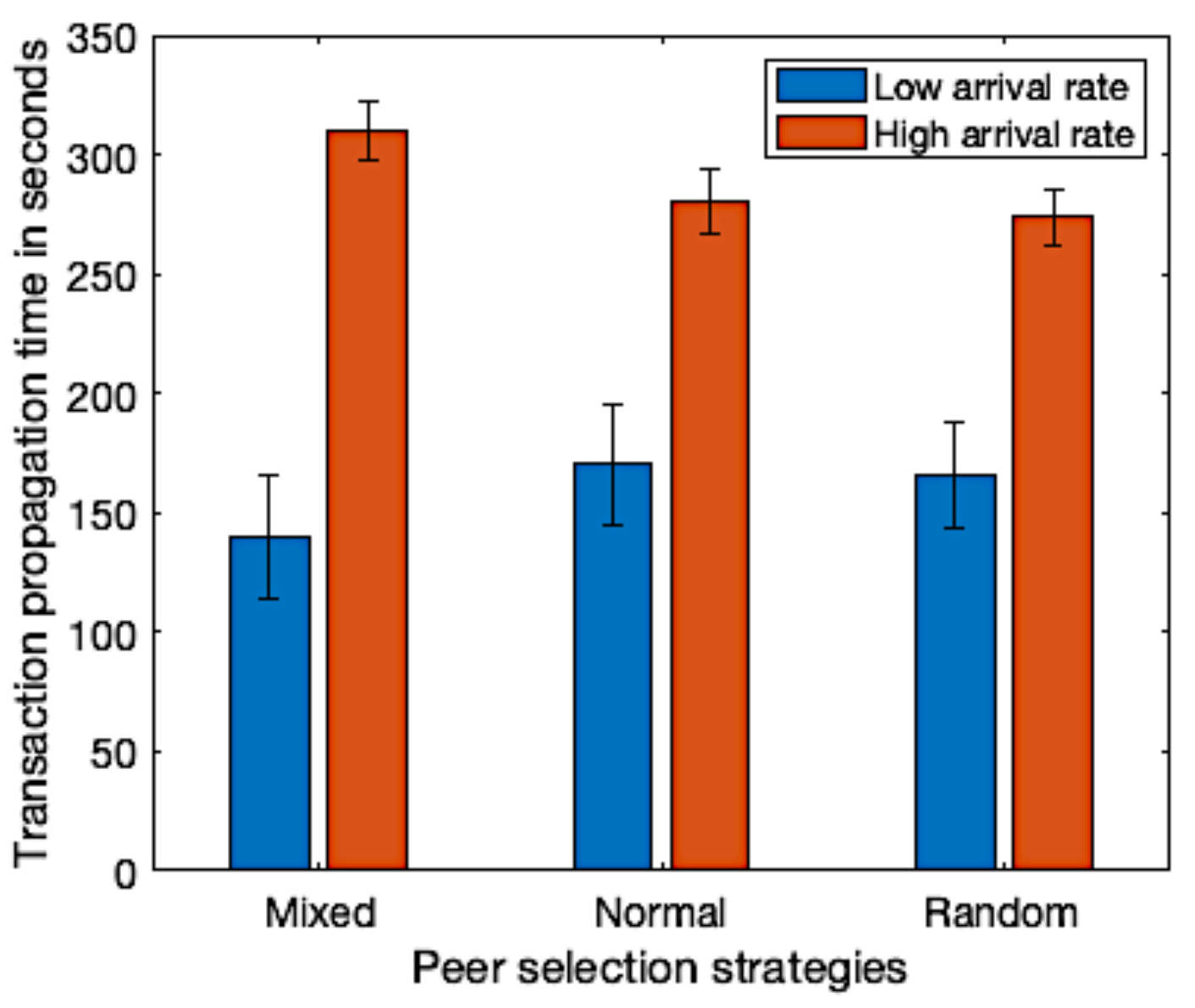}
   \label{TranProCompP4}
    }
    \label{protime}
  \caption{Average transaction propagation times for the various peer formation strategies, no. of peers $P$ and low (3 t/min) and high (6 t/min) intensity generation rate $\lambda$. Error bars indicate 95\% confidence intervals from 10 independent runs}  
\end{figure}

Fig. \ref{protime} shows the average transactions propagation time in seconds while considering different peer formation strategies. The x-axis represents peer selection strategies, the y-axis represents the propagation delay in seconds, and the legend shows the arrival intensity. 
Fig.~\ref{TranProComp} and~\ref{TranProCompP4} reports that when the arrival rate is high, which means each node generates on average six transactions per second, in respective of the number of peers per node, the transaction propagation increases. However, with a low arrival rate, three transactions per minute per node, the transaction propagation is smaller than 170 seconds. In addition, when the number of peers is higher, the normal approach tends to perform less overall random-based peer selection better than the other two. 




\subsubsection{Distribution of propagation time}

\begin{figure}[ht!]
\centering
  \includegraphics[width=0.9\linewidth, height=0.4\linewidth]{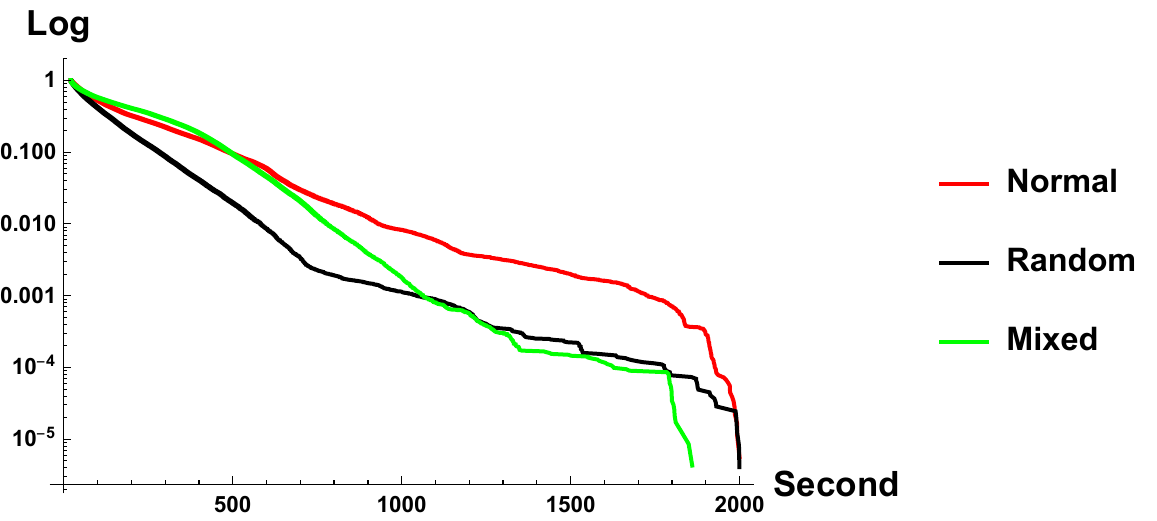}
   \caption{Transaction propagation delay, where $\lambda=3$ transactions per minute per each node (low intensity) Y-AXIS LABLE: $ P(t_a - t_g > t)$ X-AXIS LABLE: $t$}
  \label{TranPro}
\end{figure}

\begin{figure}[ht!]
\centering
  \includegraphics[width=0.95\linewidth, height=0.4\linewidth]{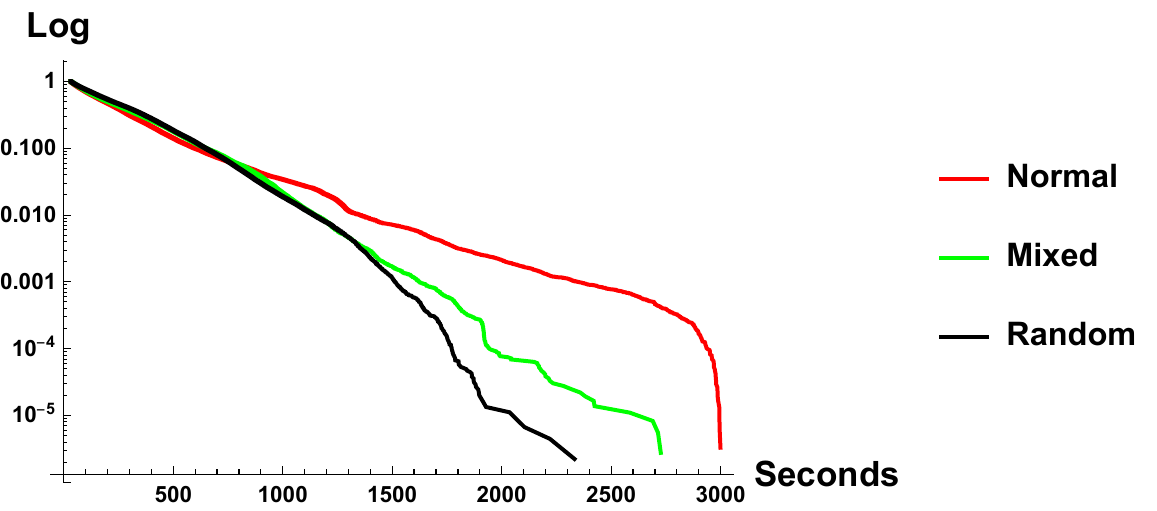}
   \caption{Transaction propagation delay, where $\lambda=6$ transactions per minute per each node (high intensity) Y-AXIS LABLE: $ P(t_a - t_g > t)$ X-AXIS LABLE: $t$}
  \label{TranProI6}
\end{figure}

Fig.~\ref{TranPro} and~\ref{TranProI6} reports the sample result showing the transaction propagation time in terms of low to a high arrival rate while the number of peers is fixed to eight. The x-axis represents the propagation time in seconds. The y-axis is the log, while the three peer formation strategies are used.

Fig.~\ref{TranPro} shows the transactions propagation delay with the three peer selection stratiegies. In most cases (80\%), the figure reports that transaction propagation in random peer selection has less than 300 seconds propagation time, whereas it has 400 seconds during the Mixed approach, while for Normal peer selection transactions sees close to 500 seconds propagation time. In three cases, the transaction propagation time can grow more than 1000 seconds in 1\% of the cases. Relatively, 90\% of the transactions see propagation time less than 500 seconds for mixed and normal approaches. Nevertheless, random-based peer formation brings less than 450 seconds of propagation time. 

 Fig.~\ref{TranProI6} also illustrates the transactions propagations delay while three peer selections are considered. In most cases (80\%), the figure reports that transaction propagation in random peer selection has less than 400 seconds propagation time, whereas it has 500 seconds during the Mixed and Normal peer selection approach. In three cases, the transaction propagation time can grow more than 1500 seconds in 1\% of the cases. Relatively, 90\% of the transactions see propagation time less than 700 seconds for mixed and normal approaches. Nevertheless, random-based peer formation brings less than 600 seconds of propagation time.

Overall, for low transaction intensity, random peer selection performs better than the other two approaches. However, when we pushed the intensity by double, at least 60\% of the time, all the peer selection strategies brought a very close transaction propagation delay.

\section{Transaction Confirmation time}\label{sec-trasconf}

In Bitcoin, the transaction is considered confirmed six blocks deep in the main chain. This ensures no double-spending while maintaining the security by linking the previous block with the other six blocks requiring more computational effort to modify the confirmed transactions.  In the 'regtest' setup, the transaction is considered valid 101 blocks deep.  However, our analysis used six blocks deep confirmation to obtain results representative for the live Bitcoin blockchain.  
Fig.~\ref{TranProConfNodeseq} demonstrates the time sequence of transaction confirmation time. The transaction confirmation time is the difference of the $t_g$ and the $t_{\rm conf}$. The $t_{\rm conf}$ is the amount of time for the Bitcoin network to generate six valid blocks. As similar to the previous case, $t_g$ is the time a transaction is generated by one of the nodes, and $t_{\rm conf}$ is the time between the blocks from the main chain extracted at node 1. As we can see from Fig.~\ref{TranProConfNodeseq} the red line indicates the time sequence of the transaction confirmation time. 

\begin{figure}[ht!]
    \centering
    \subfigure[$P=8$]
    {
    \includegraphics[width=0.45\linewidth, height=0.4\linewidth]{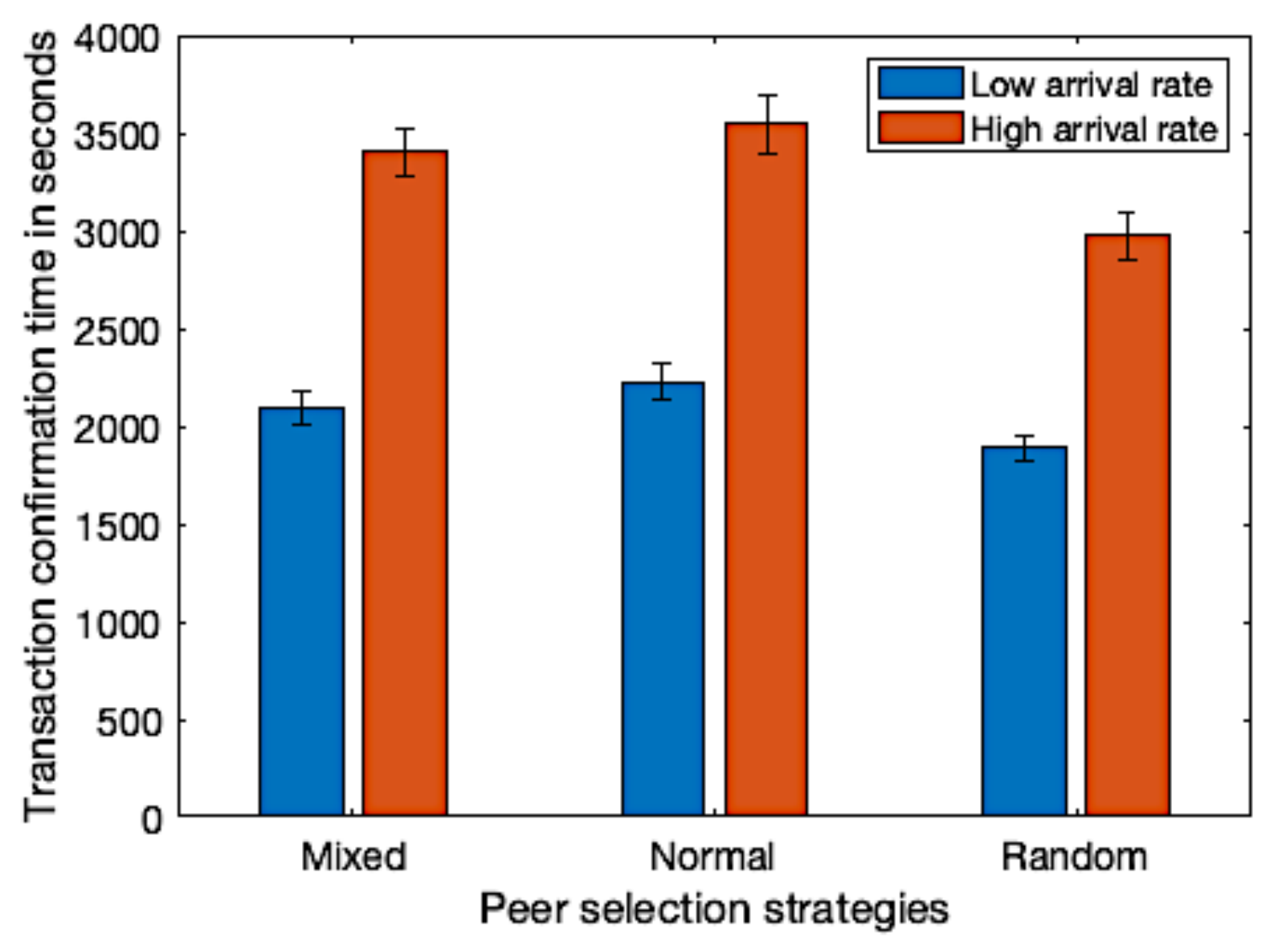}
   \label{TranConfComp}
    }
    \subfigure[$P=4$] 
    {
    \includegraphics[width=0.45\linewidth, height=0.4\linewidth]{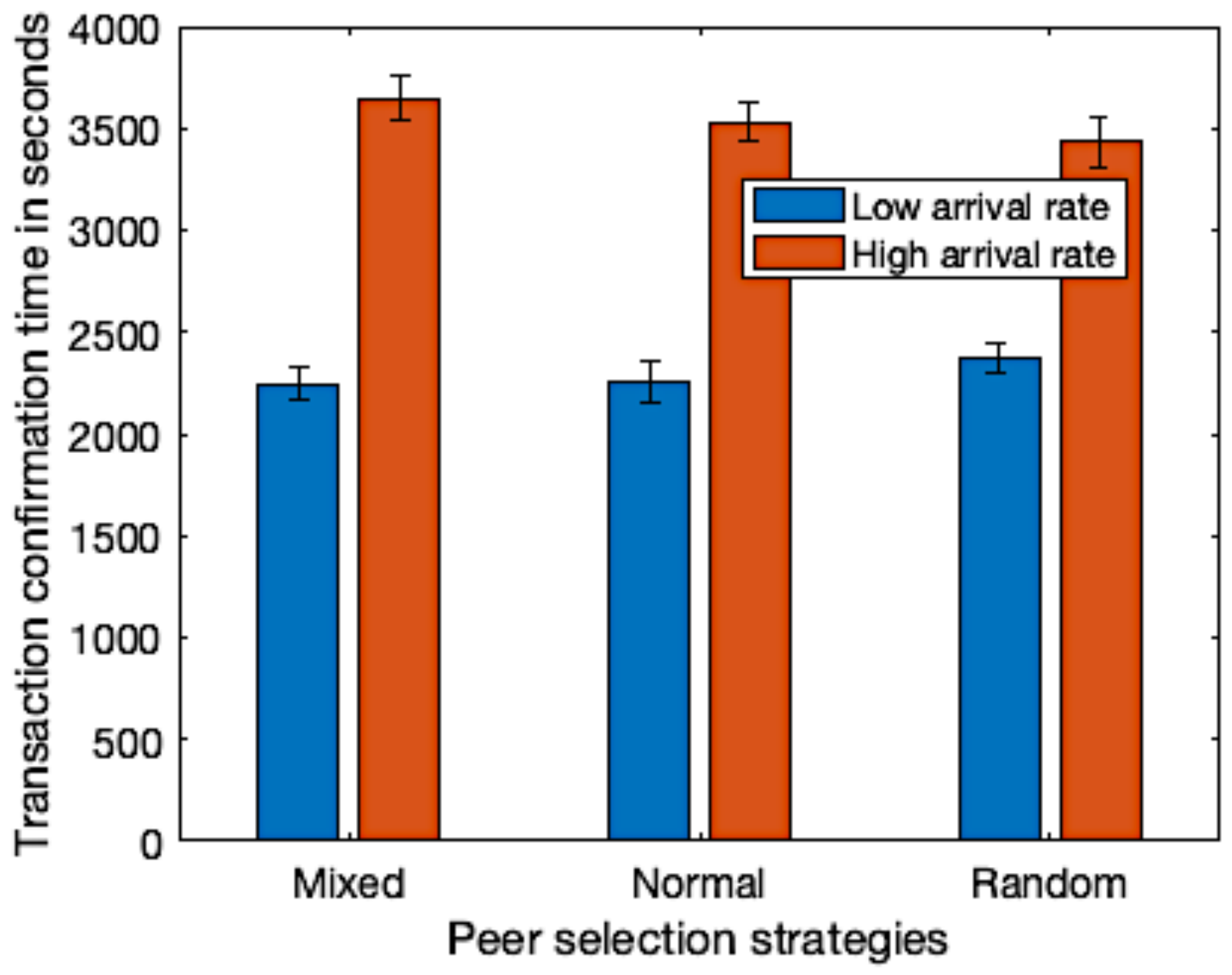}
   \label{TranConfCompP4}
    }
    \label{conftime}
  \caption{Average transaction confirmation times for the various peer formation strategies, no. of peers $P$ and low (3 t/min) and high (6 t/min) intensity generation rate $\lambda$. Error bars indicate 95\% confidence intervals from 10 independent runs}  
\end{figure}

Fig.~\ref{conftime} shows the transaction confirmation for the different peer formations strategies and number of peers. The figure also shows that peer formation strategy impacts the overall confirmation time of a transaction. The x-axis represent the peer selection strategies while the y-axis indicates the confirmation time in seconds. Fig.~\ref{TranConfComp} and~\ref{TranConfCompP4} reports the average transaction confirmation time, considering arrival rates and peer selection strategies. The plots show that in respect of the number of peers per node, the arrival rate has a higher impact on the confirmation time. It is worth highlighting that peer formation strategies bring less effect when the arrival rate is lower. 

%

\subsubsection{Distribution of the confimation time}
Fig.~\ref{TranPro} and~\ref{TranProI6}  shows the distribution of the transaction confirmation times in terms of low to a high arrival rate with the number of peers is fixed to eight. The x-axis represents the confirmation time in seconds. The y-axis is the log, while the three peer formation strategies are used.
\begin{figure}[ht!]
\centering
  \includegraphics[width=0.8\linewidth, height=0.4\linewidth]{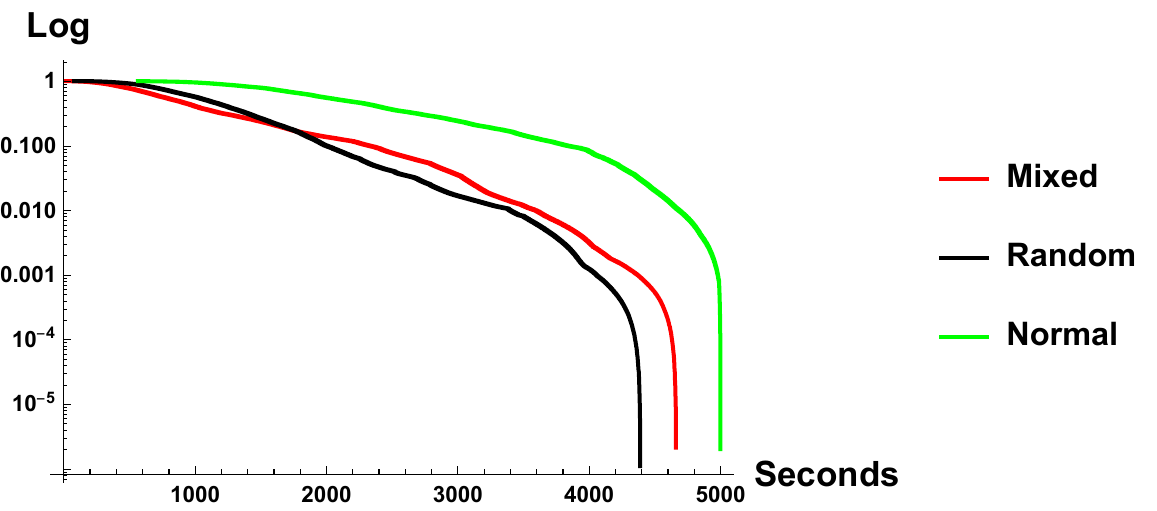}
   \caption{Transaction confirmation time, where $\lambda=3$ transactions per minute per each node (low intensity) Y-AXIS LABLE: $ P(t_g - b_{gi+6} > t)$ X-AXIS LABLE: $t$}
  \label{TranConf}
\end{figure} 

Fig.~\ref{TranConf} reports the transaction confirmation time in seconds. The x-axis represents the time it takes for a transaction to get confirmed, and the y-axis represents the log. The three peer selection strategies are compared. In almost 80\% of the cases, random and mixed peer formation strategies produced transaction confirmation time less than 1654 seconds, while the normal approach introduces twice the confirmation time. In 1\% of the time, mixed and random strategies give confirmation time greater than 2000 seconds, while normal approaches double this amount.  


\begin{figure}[ht!]
\centering
  \includegraphics[width=0.8\linewidth, height=0.4\linewidth]{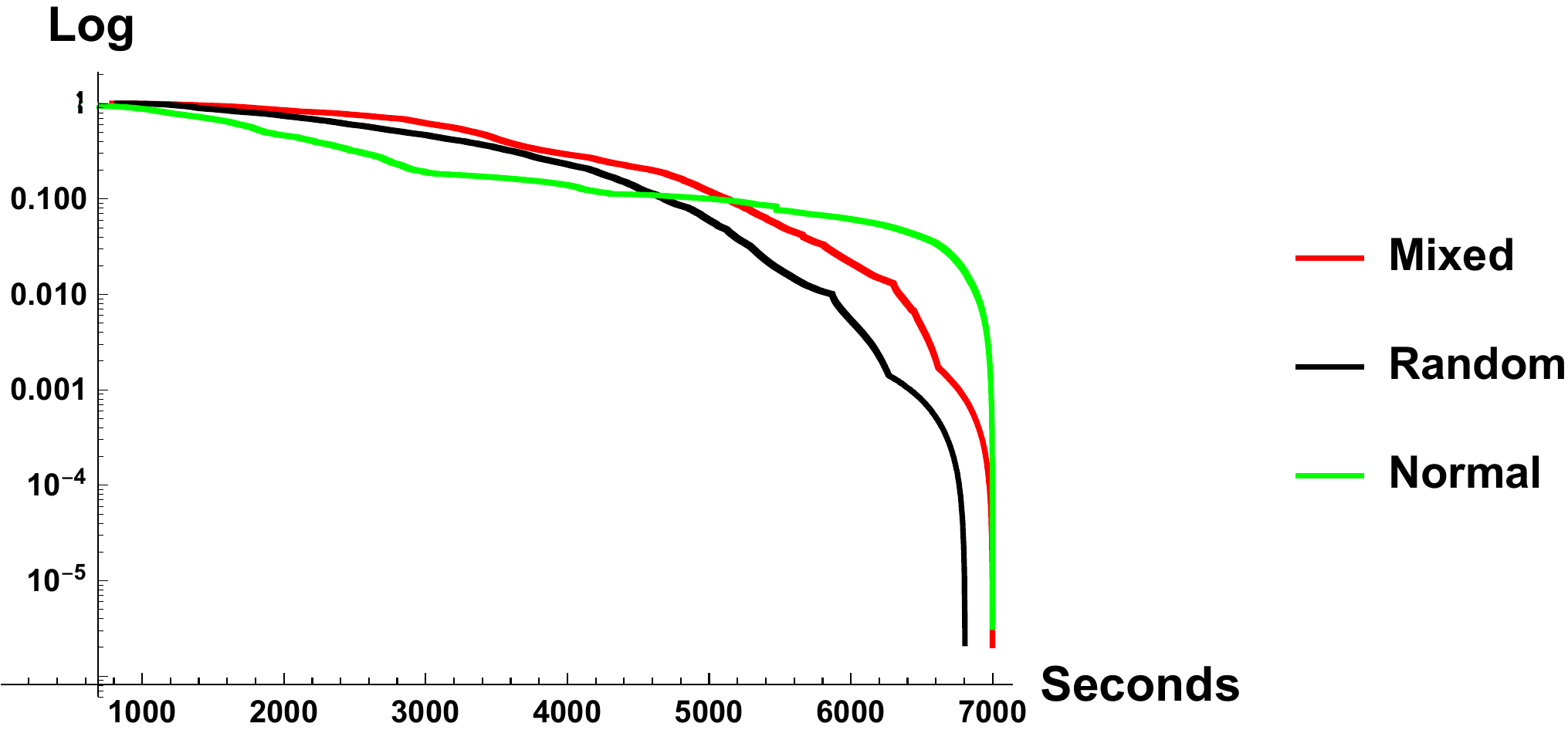}
   \caption{Transaction confirmation time, where $\lambda=6$ transactions per minute per each node (high intensity) Y-AXIS LABLE: $ P(t_g - b_{gi+6} > t)$ X-AXIS LABLE: $t$}
  \label{TranConfI6}
\end{figure} 

Fig.~\ref{TranConfI6} reports the transaction confirmation time in seconds. The x-axis represents the time for a transaction to get confirmed, and the y-axis represents the log. The three peer selection strategies are compared. In almost 80\% of the cases, random and mixed peer formation strategies produced transaction confirmation time less than 4000 seconds, while the normal approach introduced 1000 seconds less confirmation time. In 1\% of the time, all the strategies give confirmation time greater than 5000 seconds.

Overall, for low transaction intensity, random peer selection performs better than the other two approaches. However, when we pushed the intensity by double, we saw the all strategies yields more  similar distributions. Increasing the arrival intensity by double also affected the confirmation time. More transactions see higher confirmation time. 

\section{temporary forking }\label{sec-fork}

A temporary fork occurs when two miners independently find and publish a new block referencing the same previous block. The end-to-end delay between nodes inevitably leads to a temporary fork. This end-to-end delay also depends on the network topology that synchronizes between nodes. In this section, we demonstrate how the peer selection strategies affect the fork rate. 
\subsection{Introduction to temporary fork}


\begin{figure}[ht!]
\centering
  \includegraphics[width=0.9\linewidth, height=0.25\linewidth]{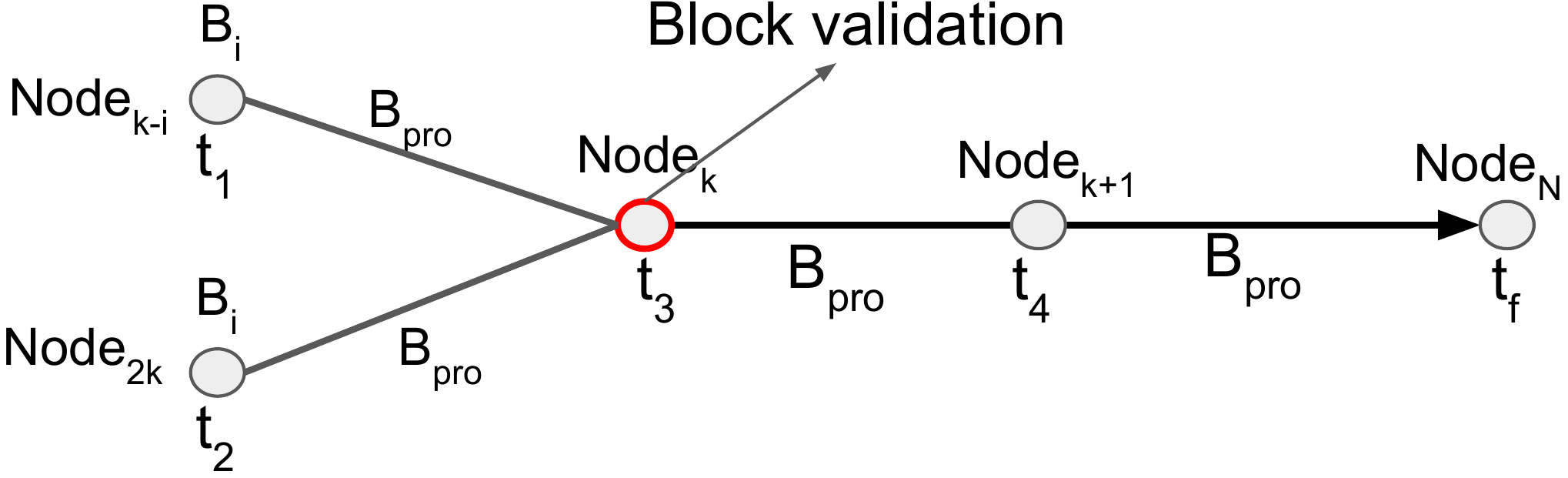}
   \caption{Block propagation time sequence }
  \label{Blockpro}
\end{figure}

Fig.~\ref{Blockpro} llustrates the time sequence of block propagation. Two blocks are generated at time $t_1$ and $t_2$ then pushed to the neighbor nodes with some $B_{pro}$ delay.  When Node$_k$ receives these two blocks simultaneously, it validates each block is valid. Suppose both blocks point to the same previous hash of the block. Then the node compares the number of confirmations and an earlier timestamp. It selects one block based on these criteria, increases the confirmation, and forwards it to the neighbor nodes. Similarly, Node$_{k+1}$ will do the same operations, and this will increase the number of confirmation numbers of the valid block that will lead the orphan (fork) block to become less important as time goes. Once all the N nodes see these two blocks, the network ignores the orphan blocks while the valid block is added to the main chain~\cite{short}. 
In this way, the Bitcoin network maintains the ledger's consistency and security. However, this temporary fork impacts the overall performance of the technology. The validated transaction from  an orphan block which are not part of the valid block goes back to mempool to wait for pick-up again, increasing the average confirmation time. In addition, miners who created the ignored block (orphan block) wasted considerable resources for less gain because of the propagation delay. 

\subsubsection{Example}

\begin{figure}[ht!]
    \centering
    \subfigure[$P=8$]
    {
    \includegraphics[width=0.45\linewidth, height=0.45\linewidth]{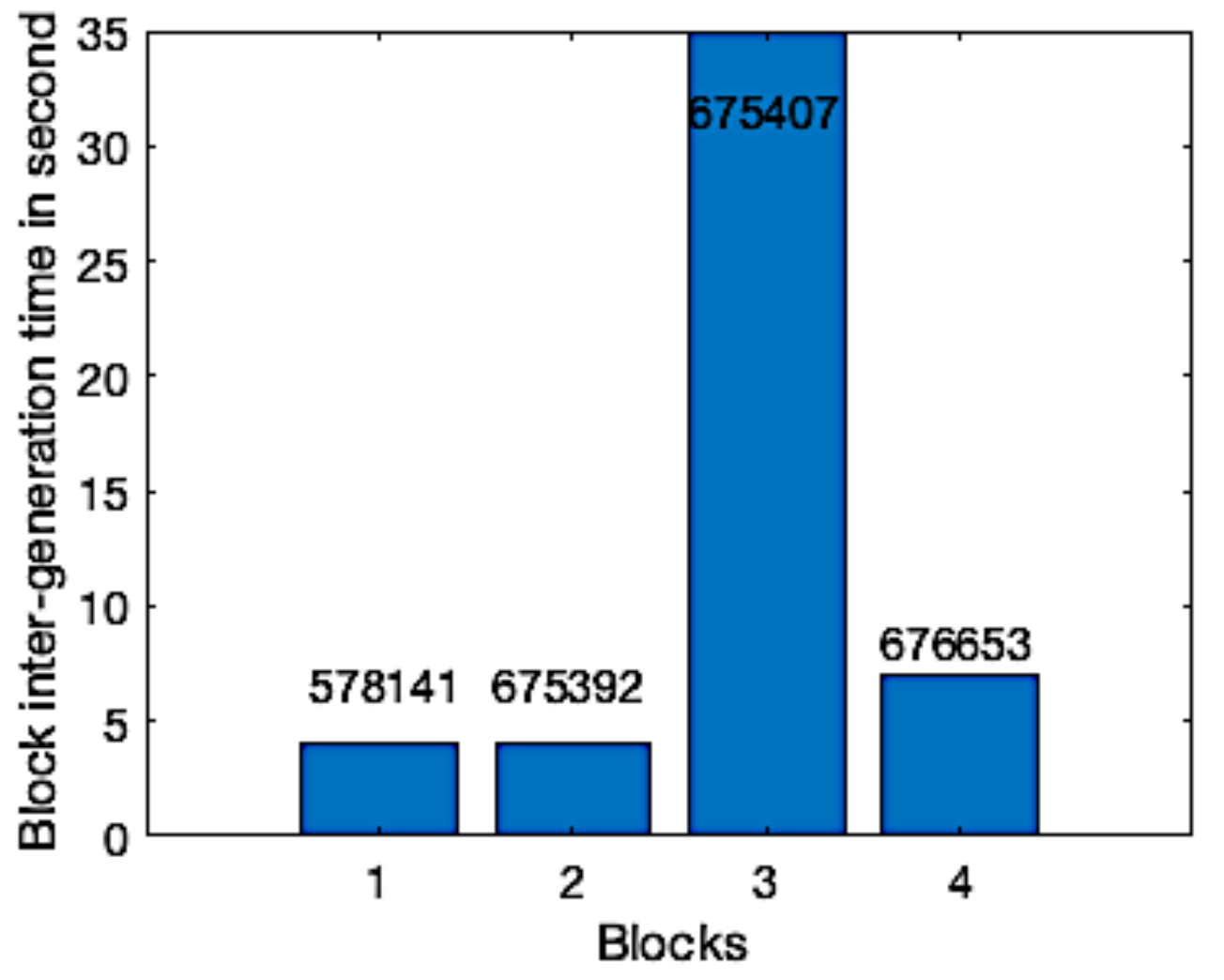}
   \label{forkmeas}
    }
    \subfigure[$P=8$] 
    {
    \includegraphics[width=0.45\linewidth, height=0.45\linewidth]{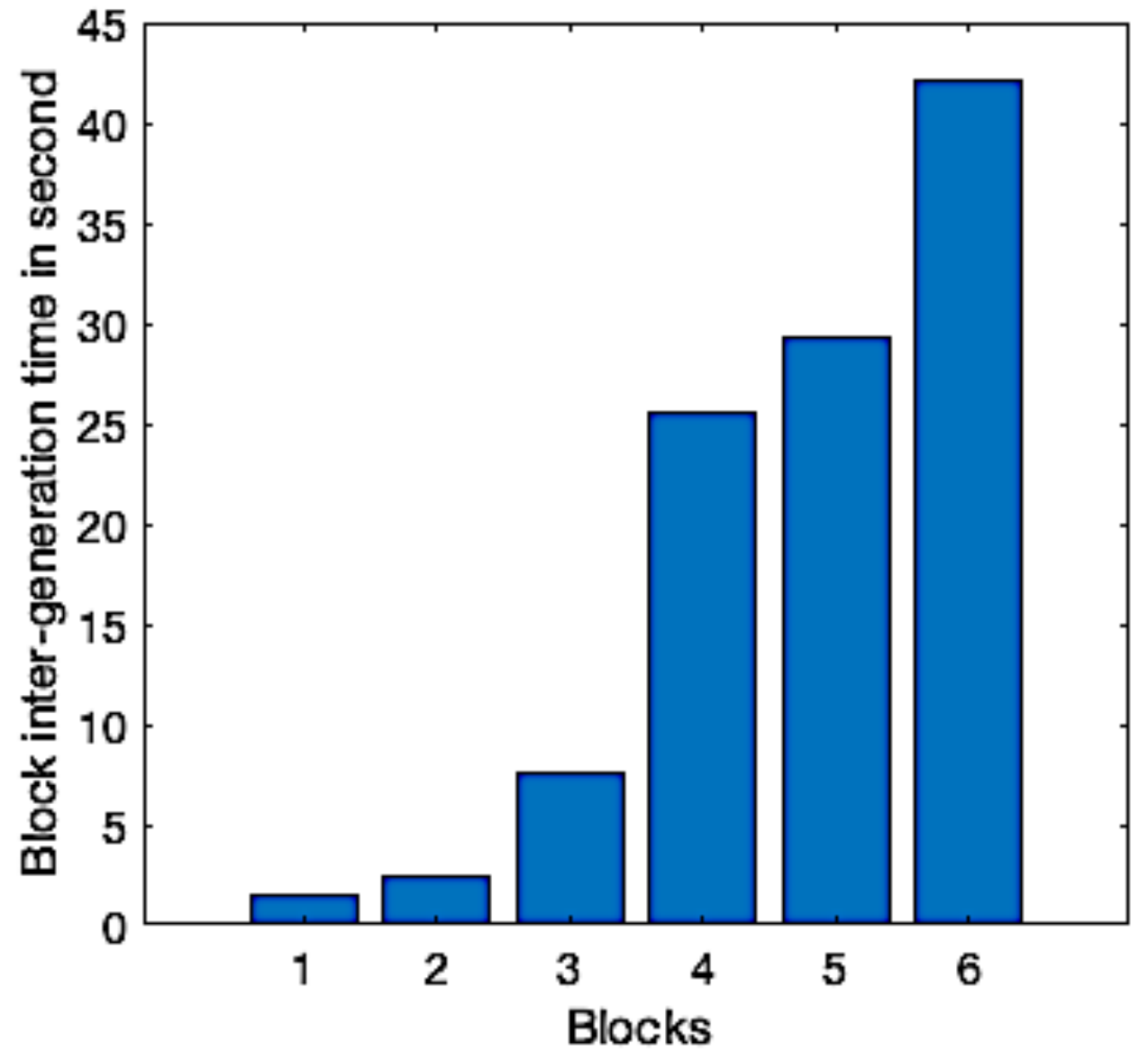}
   \label{TranConfI6P4}
    }
    \label{Fork}
  \caption{Fork vs valid block inter-generation time in seconds}  
\end{figure}

Based on our full independent Bitcoin live node~\cite{transHandling}, we were able to see four valid forks in the main chain from 578141 to 678853 block height. These four blocks contain transactions from 1200 to 2400 per block. The average  generation time between to blocks forming a fork is 12.5 seconds. Fig.~\ref{Fork} reports the inter-block generation time between fork and valid block in the Bitcoin network. The x-axis represents the blocks where the fork happened, and the y-axis indicates the inter-block generation time in seconds between fork and valid block. As we can see from the figure, the maximum inter-block generation time between valid and fork block is 35 seconds, which happened in the 675407 block height.

Fig.~\ref{TranConfI6P4} shows the block inter-generation time between valid and fork block, while the normal based peer selection, high arrival rate, and pl=8. The figure illustrates that block inter-generated time less than 40 seconds has a high probability of becoming fork.  This plot is to demonstrate what we see from Fig.~\ref{forkmeas}, which is from live bitcoin node, is also seen from our setup.


\subsection{Peer selection strategies impact on the fork rate}

The peer selection strategies impact the performance of the technology. In this work, we considered three strategies and showed the effect on the transaction propagation and confirmation time. In this subsection, we demonstrate its impact on the occurrence of forks. 


\begin{figure}[ht!]
    \centering
    \subfigure[Number of fork]
    {
    \includegraphics[width=0.45\linewidth, height=0.5\linewidth]{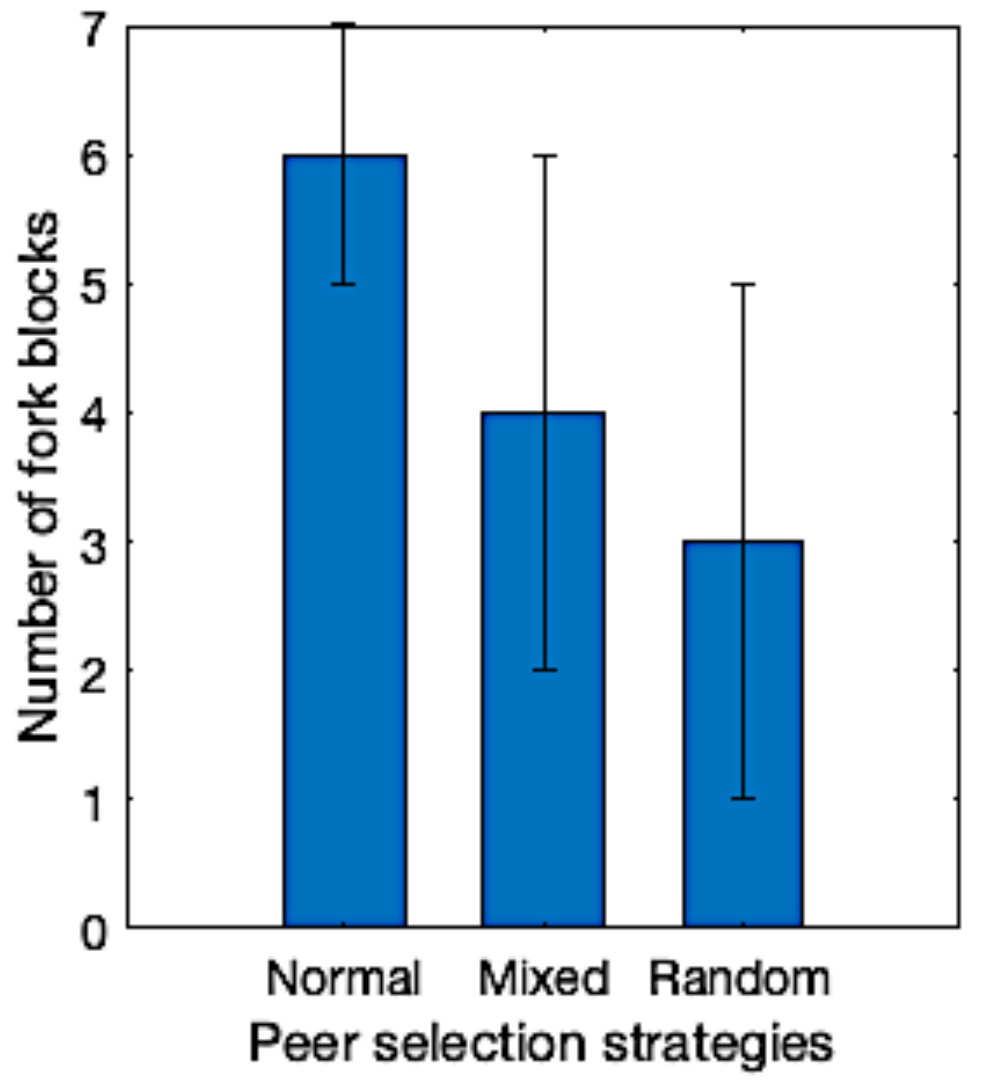}
   \label{forkpeer}
    }
    \subfigure[Confirmation time] 
    {
    \includegraphics[width=0.45\linewidth, height=0.5\linewidth]{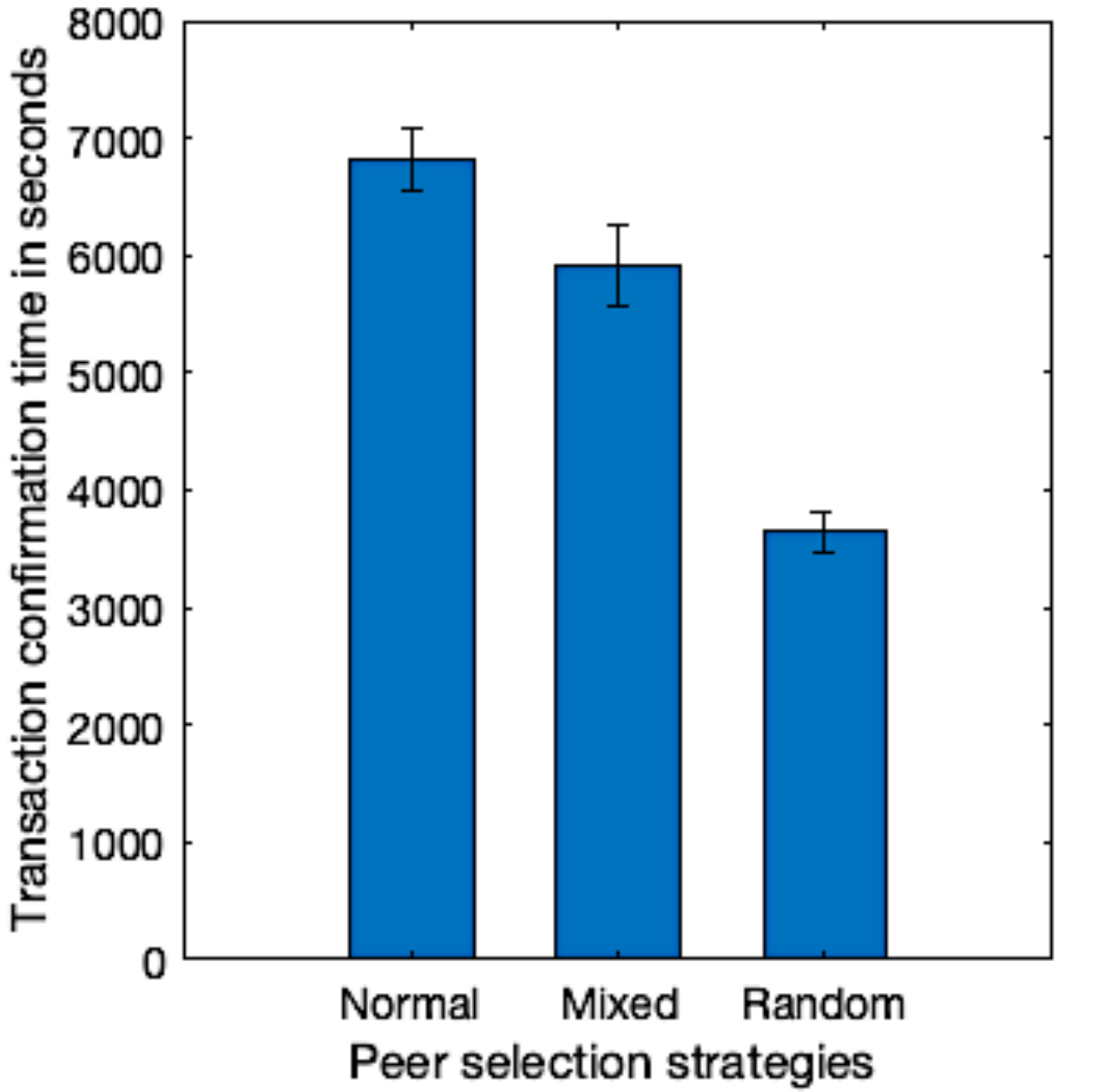}
   \label{TranConffork}
    }
    \label{ForkRes}
  \caption{Impact of fork in transaction confirmation time}  
\end{figure}

Fig.~\ref{forkpeer} reports the number of fork blocks happening while different peer selection strategies adopt, high arrival intensity, and the number of peers per node is 8. As we can see from the figure, the normal peer selection strategies bring a higher number of temporary forks. The mixed and random-based peer selection strategies produce less number of fork blocks. 


\subsection{The occurrence of Fork impacting transaction confirmation time }
In the fork event, transactions inside a fork block return to the mempool for pick up. This makes these transactions wait a longer time before confirmation. Fig.~\ref{TranConffork} reports the average transaction confirmation time seen by transactions inside the fork block. When the network ignores the fork block, all the transactions inside return back to the mempool for the pickup, the main issue with this approach is fork block may wait for more than one block to be ignored by the network, depending on the length of the pruned branch, which leads to the transaction stuck and waiting for a longer time. Fig.~\ref{TranConffork} also demonstrates this property, as we can see from the figure, the normal peer selection strategy produces a high number of fork blocks, which also leads to transactions waiting longer than 6000 seconds. 

Similarly, the mixed peer selection strategy also produces a close number of fork blocks as a normal peer selection approach, in which we can see the impact on the transaction confirmation time is more than 5500 seconds. However, the random peer selection strategy performs better than the first two approaches regarding the number of forks and confirmation time, leading to a transaction confirmation time of fewer than 3670 seconds. 

\subsection{Valid vs fork block correlation}
When two blocks arrive in a short time in deference while having the same hash of the previous block, we call it a temporary fork. When this event happens, one of the blocks will become part of the chain, and the other will become an orphan block. Since the comparison is based on the previous block's hash, we further analyze the extent to which valid, and fork blocks share the same transactions. Table~\ref{corr} shows the correlation in percentage between the valid and fork block while considering different peer selection strategies. 

\begin{table}[ht]
\centering
\caption{Correlation }
\begin{tabular}{|l|l|l|}
\hline
 Peer selection strategies & Correlation valid vs fork block ($(\mu, \sigma)$) \\\hline
 
Normal & (88\%, 6\%)  \\ \hline

Mixed & (90\%, 5\%)     \\ \hline

Random & (90\%, 3\%)  \\ \hline

\end{tabular}

\label{corr}
\end{table}

The mixed and normal peer selection strategies produce four to six fork blocks, which has from 90\% of the transaction are the same, but the rest 10\% are unique transactions which be forced to return to the backlog for more waiting time. Similarly, for the random-based strategy, the valid and fork block shares 90\% of the transactions are similar, but the remaining wait more time to be added to the chain. 

\section{Discussion}\label{sec-dis}
\subsubsection{Proposed approaches}
The P2P formation strategies are essential in propagating information between participating nodes. In this work, we showed that peer selection strategies affect the overall performance of Bitcoin. There has been some research work proposing schemes and methods in reducing the propagation delay in Bitcoin. This proposal focuses on either introducing a compact block~\cite{compactrelay}\cite{compact2} or having some relay nodes~\cite{relay}\cite{RelayNetwork} in the middle to provide a pipeline to push more updates to the other nodes. The compact block may introduce better performance in propagating the information based on the available bandwidth between participating nodes~\cite{compactrelay}. However, this method focuses on how to propagate blocks in the network than how to propagate transactions. Based on our observation, nodes may have a different number of arrival at the backlog waiting. The compact block method has to push more than half of the block content in such cases. 

Using relay nodes to reduce propagation delay is another method proposed by researchers. This method relies on the relay nodes having a higher number of peer nodes from the network, enabling push more updates in the network. The main challenge in this approach is that the relay nodes become a security bridge or vulnerability point. Attacking these nodes or taking control gives extra incentive to earn more or disrupt the overall activity in the network.  

The best strategy to improve the propagation time is to improve the communication protocol. The protocol spends significant time validating and updating the same transaction. Furthermore, reducing the peer to peer network diameter by having peers other than the nearest may improve. The  random strategies in this paper are simple examples

\subsubsection{Transaction propagation and confirmation times}
The transaction propagation and confirmation time show some values higher than expected. This is because of the impact of the P2P formation strategies and P2P legacy relaying protocol. Some of the transactions have to return to mempool because of fork occurrence. For instance, the normal-based approach produces more forks than the other two approaches. In such cases, the transaction inside the fork block returns to the backlog for pick up, some of these transactions will be added to the new recent block, but others will wait for another block generation event. In addition to this, the processing capacity of the raspberry pi devices may contribute to some extent. Although we analyzed to observe the total usage, the bitcoin, on average, in each device uses 114\% CPU and 16\% RAM. It is worth highlighting that the raspberry pi used has 64 bit quad-core Cortex-A53 and Cortex-A72, which is good enough to handle the traffic generated from Bitcoin and background processing.  
\subsubsection{Impact of temporary fork}
The number of fork event occurrences has been reduced recently with the new Bitcoin core release~\cite{short}. However, the Bitcoin network is still not tested if it can handle high loads. Based on the current state where 3.3 to 7.2 transactions are processed per second, having arrivals at the mempool from 1700-2600 transactions waiting for pick up~\cite{Bitaps, Blockstream, Explorer, Btc}. The P2P network may handle processing and propagating updates with some acceptable performance index. However, when we pushed the load to 5500 to 6000 transactions at the backlog, the performance reduced significantly from propagating transactions in 10 seconds into 250-350 seconds. It also impacted the number of fork block occurrences in the network, making some transactions wait more than the expected confirmation time. For instance, for distance-based peer formation strategies, the number of prude branches is higher because each node validates new arrivals before propagating to the neighbor nodes. In such cases, more delays happen in the network than having few random peer links. This shows that the P2P network protocol requires significant improvement and research to improve its capacity. 
\section{Conclusion} \label{sec-con}
In this paper, we analyzed the impact of peer formation strategies, arrival rate, and the number of peers on the overall performance of the technology. Specifically, we developed a testbed to mimic the Bitcoin P2P network, which enabled us to conduct deep insight into the traffic characteristics of the overlay network. The analysis shows that the transaction validation and propagation takes longer than expected, even with a low arrival rate and a high number of connected nodes. In addition, the peer formation strategy adopted by the bitcoin community is highly reliable in finding peers with low latency response, which makes it less efficient since each node has to validate each transaction and block before forwarding to the next peer. Considering a few random peer node selections improved the performance, this can improve the peer formation strategies adopted by the nodes. These results show that distance-based peer formation alone may not bring optimal solutions. Performing an independent investigation on improving the P2P communication protocol may enhance the reduction of validating the same transactions/block multiple times. 

\AtNextBibliography{\footnotesize}
{\footnotesize \printbibliography}

\end{document}